\input harvmac
\input amssym


\def\CN{{\cal N}}
\def\CL{{\cal L}}
\def\CV{{\cal V}}

\def\BQ{{\overline Q}}
\def\BD{{\overline D}}
\def\Bb{{\overline b}}

\def\BF{{\overline F}}
\def\BW{{\overline W}}
\def\BG{{\overline G}}
\def\BE{{\overline E}}
\def\BJ{{\overline J}}
\def\BT{{\overline T}}
\def\Bphi{{\overline \phi}}
\def\Bpsi{{\overline \psi}}
\def\Bxi{{\overline \xi}}
\def\Bchi{{\overline \chi}}
\def\Blambda{{\overline \lambda}}
\def\Bsigma{{\overline \sigma}}
\def\Bepsilon{{\overline \epsilon}}
\def\BPhi{{\overline \Phi}}
\def\BPsi{{\overline \Psi}}
\def\BXi{{\overline \Xi}}
\def\Btheta{{\overline \theta}}
\def\BLambda{{\overline \Lambda}}
\def\BSigma{{\overline \Sigma}}
\def\BUpsilon{{\overline \Upsilon}}
\def\BCL{{\overline \CL}}
\def\BCD{{\overline \CD}}
\def\BCV{{\overline \CV}}

\def\Tr{{\rm Tr}}

\Title{\vbox{\baselineskip 12pt \hbox{KUNS-1955}\hbox{\tt hep-th/0503074}}}
{Gauged Linear Sigma Model on Supermanifold}

\centerline{Shigenori Seki\footnote{$^\dagger$}
{\tt seki@gauge.scphys.kyoto-u.ac.jp}}
\medskip\centerline{\it Department of Physics, Graduate School of Science}
\centerline{\it Kyoto University, Kyoto 606-8502, Japan}
\bigskip\centerline{and}\bigskip
\centerline{Katsuyuki Sugiyama\footnote{$^\ddagger$}
{\tt sugiyama@phys.h.kyoto-u.ac.jp}}
\bigskip\centerline{\it Department of Physics, Graduate School of Science, Yoshida South}
\centerline{\it Kyoto University, Kyoto 606-8501, Japan}

\vskip .3in

\centerline{{\bf abstract}}

We formulate a gauged linear sigma model on a supermanifold. 
The structure of classical vacua is investigated and 
one-loop corrections are calculated for an auxiliary D-field. 
We find out a constraint 
for the one-loop divergence to vanish, which is consistent with a Ricci flatness
condition for the supermanifold. 
Two types of D-branes are obtained 
by analysing supersymmetric boundary conditions.
We also provide extensions to a non-abelian gauged linear sigma model
and a $(0,2)$ supersymmetric model.

\Date{March 2005}

\lref\HIV{K.~Hori, A.~Iqbal and C.~Vafa, ``D-Branes and Mirror Symmetry,'' {\tt hep-th/0005247}.}
\lref\HV{K.~Hori and C.~Vafa, ``Mirror Symmetry,'' {\tt hep-th/0002222}.}
\lref\MP{D.~R.~Morrison and M.~R.~Plesser, ``Summing the Instantons: Quantum Cohomology and Mirror Symmetry in Toric Varieties,'' Nucl. Phys. B440 (1995) 279, {\tt hep-th/9412236}.}
\lref\OOY{H.~Ooguri, Y.~Oz and Z.~Yin, ``D-Branes on Calabi-Yau Spaces and Their Mirrors,'' Nucl. Phys. B477 (1996) 407, {\tt hep-th/9606112}.}
\lref\Wii{E.~Witten, ``Phases of $N=2$ Theories In Two Dimensions,'' Nucl. Phys. B403 (1993) 159, {\tt hep-th/9301042}.}
\lref\Wiii{E.~Witten, ``The Verlinde Algebra and The Cohomology of The Grassmannian,'' {\tt hep-th/9312104}.}
\lref\WB{J.~Wess and J.~Bagger, ``Supersymmetry and Supergravity,'' Princeton University Press (1976).}
\lref\Ah{C.~Ahn, ``Mirror Symmetry of Calabi-Yau Supermanifolds,'' {\tt hep-th/0407009}.}
\lref\Ahi{C.~Ahn, ``${\cal N}=1$ Conformal Supergravity and Twistor-String Theory,'' JHEP 0410 (2004) 064, {\tt hep-th/0409195}.}
\lref\Ahii{C.~Ahn, ``$\CN =2$ Conformal Supergravity from Twistor-String Theory,'' {\tt hep-th/0412202}.}
\lref\AV{M.~Aganagic and C.~Vafa, ``Mirror Symmetry and Supermanifolds,'' {\tt hep-th/0403192}.}
\lref\Be{N.~Berkovits, ``An Alternative String Theory in Twistor Space for N=4 Super-Yang-Mills,'' Phys. Rev. Lett. 93 (2004) 011601, {\tt hep-th/0402045}.}
\lref\BBDD{S.~J.~Bidder, N.~E.~J.~Bjerrum-Bohr, L.~J.~Dixon and D.~C.~Dunbar, ``$N=1$ Supersymmetric One-loop Amplitudes and the Holomorphic Anomaly of Unitarity Cuts,'' {\tt hep-th/0410296}.}
\lref\BBK{I.~Bena, Z.~Bern and D.~A.~Kosower, ``Twistor-Space Recursive Formulation of Gauge-Theory Amplitudes,'' {\tt hep-th/0406133}.}
\lref\BBKR{I.~Bena, Z.~Bern, D.~A.~Kosower and R.~Roiban, ``Loops in Twistor Space,'' {\tt hep-th/0410054}.}
\lref\BBST{J.~Bedford, A.~Brandhuber, B.~Spence and G.~Travaglini, ``A Twistor Approach to One-Loop Amplitudes in $\CN =1$ Supersymmetric Yang-Mills Theory,'' {\tt hep-th/0410280}.}
\lref\BBSTi{J.~Bedford, A.~Brandhuber, B.~Spence and G.~Travaglini, ``Non-Supersymmetric Loop Amplitudes and MHV Vertices,'' {\tt hep-th/0412108}.}
\lref\BCF{R.~Britto, F.~Cachazo and B.~Feng, ``Computing One-Loop Amplitudes from the Holomorphic Anomaly of Unitarity Cuts,'' {\tt hep-th/0410179}.}
\lref\BCFi{R.~Britto, F.~Cachazo and B.~Feng, ``Coplanarity in Twistor Space of ${\cal N}=4$ Next-to-MHV One-Loop Amplitude Coefficients,'' {\tt hep-th/0411107}.}
\lref\BCFii{R.~Britto, F.~Cachazo and B.~Feng, ``Generalized Unitarity and One-Loop Amplitudes in $\CN = 4$ Super-Yang-Mills,'' {\tt hep-th/0412103}.}
\lref\BDDK{Z.~Bern, V.~D.~Duca, L.~J.~Dixon and D.~A.~Kosower, ``All Non-Maximally-Helicity-Violating One-Loop Seven-Gluon Amplitudes in $\CN =4$ Super-Yang-Mills Theory,'' {\tt hep-th/0410224}.}
\lref\BDK{Z.~Bern, L.~J.~Dixon and D.~A.~Kosower, ``All Next-to-Maximally-Helicity-Violating One-Loop Gluon Amplitudes in $\CN =4$ Super-Yang-Mills Theory,'' {\tt hep-th/0412210}.}
\lref\BDRSS{A.~Belhaj, L.~B.~Drissi, J.~Rasmussen, E.~H.~Saidi and A.~Sebbar, ``Toric Calabi-Yau Supermanifolds and Mirror Symmetry,'' {\tt hep-th/0410291}.}
\lref\BM{N.~Berkovits and L.~Motl, ``Cubic Twistorial String Field Theory,'' JHEP 0404 (2004) 056, {\tt hep-th/0403187}.}
\lref\BST{A.~Brandhuber, B.~Spence and G.~Travaglini, ``One-Loop Gauge Theory Amplitudes in ${\cal N}=4$ Super Yang-Mills from MHV Vertices,'' {\tt hep-th/0407214}.}
\lref\BWi{N.~Berkovits and E.~Witten, ``Conformal Supergravity in Twistor-String Theory,'' JHEP 0408 (2004) 009, {\tt hep-th/0406051}.}
\lref\Ca{F.~Cachazo, ``Holomorphic Anomaly of Unitarity Cuts and One-Loop Gauge Theory Amplitudes,'' {\tt hep-th/0410077}.}
\lref\CSW{F.~Cachazo, P.~Svrcek and E.~Witten, ``MHV Vertices and Tree Amplitudes in Gauge Theory,'' JHEP 0409 (2004) 006, {\tt hep-th/0403047}.}
\lref\CSWi{F.~Cachazo, P.~Svrcek and E.~Witten, ``Twistor Space Structure of One-Loop Amplitudes in Gauge Theory,'' JHEP 0410 (2004) 074, {\tt hep-th/0406177}.}
\lref\CSWii{F.~Cachazo, P.~Svrcek and E.~Witten, ``Gauge Theory Amplitudes in Twistor Space and Holomorphic Anomaly,'' JHEP 0410 (2004) 077, {\tt hep-th/0409245}.}
\lref\DGK{L.~J.~Dixon, E.~W.~N.~Glover and V.~V.~Khoze, ``MHV Rules for Higgs Plus Multi-Gluon Amplitudes,'' {\tt hep-th/0411092}.}
\lref\GGK{G.~Georgiou, E.~W.~N.~Glover and V.~V.~Khoze, ``Non-MHV Tree Amplitudes in Gauge Theory,'' JHEP 0407 (2004) 048, {\tt hep-th/0407027}.}
\lref\GJS{S.~Govindarajan, T.~Jayaraman and T.~Sarkar, ``Worldsheet Approaches to D-branes on Supersymmetric Cycles,'' Nucl. Phys. B580 (2000) 519, {\tt hep-th/9907131}.}
\lref\GJSi{S.~Govindarajan, T.~Jayaraman and T.~Sarkar, ``On D-branes from Gauged Linear Sigma Models,'' Nucl. Phys. B593 (2001) 155, {\tt hep-th/0007075}.}
\lref\GK{G.~Georgiou and V.~V.~Khoze, ``Tree Amplitudes in Gauge Theory as Scalar MHV Diagrams,'' JHEP 0405 (2004) 070, {\tt hep-th/0404072}.}
\lref\GKRRTZ{S.~Giombi, M.~Kulaxizi, R.~Ricci, D.~Robles-Llana, D.~Trancanelli and K.~Zoubos, ``Orbifolding the Twistor String,'' {\tt hep-th/0411171}.}
\lref\GMN{S.~Gukov, L.~Motl and A.~Neitzke, ``Equivalence of Twistor Prescriptions for Super Yang-Mills,'' {\tt hep-th/0404085}.}
\lref\GP{P.~A.~Grassi and G.~Policastro, ``Super-Chern-Simons Theory as Superstring Theory,'' {\tt hep-th/0412272}.}
\lref\Ko{D.~A.~Kosower, ``Next-to-Maximal Helicity Violating Amplitudes in Gauge Theory,'' {\tt hep-th/0406175}.}
\lref\KP{S.~P.~Kumar and G.~Policastro, ``Strings in Twistor Superspace and Mirror Symmetry,'' {\tt hep-th/0405236}.}
\lref\LP{O.~Lechtenfeld and A.~D.~Popov, ``Supertwistors and Cubic String Field Theory for Open $N=2$ Strings,'' Phys. Lett. B598 (2004) 113, {\tt hep-th/0406179}.}
\lref\LW{M.~Luo and C.~Wen, ``One-Loop Maximal Helicity Violating Amplitudes in $N=4$ Super Yang-Mills Theories,'' {\tt hep-th/0410045}.}
\lref\Mo{M.~Movshev, ``Yang-Mills Theory and a Superquadric,'' {\tt hep-th/0411111}.}
\lref\NOV{N.~Nekrasov, H.~Ooguri and C.~Vafa, ``S-duality and Topological Strings,'' JHEP 0410 (2004) 009, {\tt hep-th/0403167}.}
\lref\NV{A.~Neitzke and C.~Vafa, ``${\cal N}=2$ Strings and the Twistorial Calabi-Yau,'' {\tt hep-th/0402128}.}
\lref\PR{J.~Park and S.-J.~Rey, ``Supertwistor Orbifolds: Gauge Theory Amplitudes and Topological Strings,'' {\tt hep-th/0411123}.}
\lref\PS{A.~D.~Popov and C.~Saemann, ``On Supertwistors, the Penrose-Ward Transform and ${\cal N}=4$ super Yang-Mills Theory,'' {\tt hep-th/0405123}.}
\lref\PW{A.~D.~Popov and M.~Wolf, ``Topological B-Model on Weighted Projective Spaces and Self-Dual Models in Four Dimensions,'' JHEP 0409 (2004) 007, {\tt hep-th/0406224}.}
\lref\QR{C.~Quigley and M.~Rozali, ``One-Loop MHV Amplitudes in Supersymmetric Gauge Theories,'' {\tt hep-th/0410278}.}
\lref\RSV{R.~Roiban, M.~Spradlin and A.~Volovich, ``A Googly Amplitude from the B-model in Twistor Space,'' JHEP 0404 (2004) 012, {\tt hep-th/0402016}.}
\lref\RSVi{R.~Roiban, M.~Spradlin and A.~Volovich, ``On the Tree-Level S-Matrix of Yang-Mills Theory,'' Phys. Rev. D70 (2004) 026009, {\tt hep-th/0403190}.}
\lref\RV{R.~Roiban and A.~Volovich, ``All Googly Amplitudes from the B-model in Twistor Space,'' {\tt hep-th/0402121}.}
\lref\RW{M.~Rocek and N.~Wadhwa, ``On Calabi-Yau Supermanifolds,'' {\tt hep-th/0408188}.}
\lref\Se{S.~Sethi, ``Supermanifolds, Rigid Manifolds and Mirror Symmetry,'' Nucl. Phys. B430 (1994) 31, {\tt hep-th/9404186}.}
\lref\Si{W.~Siegel, ``Untwisting the Twistor Superstring,'' {\tt hep-th/0404255}.}
\lref\SV{A.~Sinkovics and E.~Verlinde, ``A Six Dimensional View on Twistors,'' {\tt hep-th/0410014}.}
\lref\SW{X.~Su and J.-B.~Wu, ``Six-Quark Amplitudes from Fermionic MHV Vertices,'' {\tt hep-th/0409228}.}
\lref\Wi{E.~Witten, ``Perturbative Gauge Theory as a String Theory in Twistor Space,'' {\tt hep-th/0312171}.}
\lref\Wo{M.~Wolf, ``On Hidden Symmetries of a Super Gauge Theory and Twistor String Theory,'' {\tt hep-th/0412163}.}
\lref\WZ{J.-B.~Wu and C.-J.~Zhu, ``MHV Vertices and Scattering Amplitudes in Gauge Theory,'' JHEP 0407 (2004) 032, {\tt hep-th/0406085}.}
\lref\WZi{J.-B.~Wu and C.-J.~Zhu, ``MHV Vertices and Fermionic Scattering Amplitudes in Gauge Theory with Quarks and Gluinos,'' JHEP 0409 (2004) 063, {\tt hep-th/0406146}.}
\lref\Zho{C.~Zhou, ``On Ricci Flat Supermanifolds,'' {\tt hep-th/0410047}.}
\lref\Zhu{C.-J.~Zhu, ``The Googly Amplitudes in Gauge Theory,'' JHEP 0404 (2004) 032, {\tt hep-th/0403115}.}

\newsec{Introduction}

It has recently been pointed out in \Wi\ that the topological string 
theory on a twistor space 
is associated with an ${\cal N}=4$ super Yang-Mills theory.
The Maximally-Helicity-Violating (MHV) amplitude of the four dimensional 
${\cal N}=4$ super Yang-Mills theory was computed in terms of 
the topological B-model on a supermanifold $\Bbb{CP}^{3|4}$ by \Wi. 
This surprising correspondence attracts a lot of attention 
and various works have been done in this topic. 
The MHV amplitudes were analysed in \refs{\CSW\GK\WZ\WZi\LW{--}\QR} and 
non-MHV amplitudes and googly amplitudes were considered in 
\refs{\RSV\RV\Zhu\RSVi\BBK\GGK{--}\BDDK}. 
Some further investigations of the gauge theory amplitudes appeared in 
\refs{\Ko\CSWi\BST\SW\CSWii\BBKR\Ca\BCF\BBST\BBDD\DGK\BCFi\PR\BCFii\BBSTi\Wo{--}\BDK}. 
Such interesting relations between the super Yang-Mills theory 
and the topological string gave rise 
to several works \refs{\GMN\PS\LP\PW\Si\SV{--}\Mo}, 
while the string theory itself on a twistor space was developed by 
\refs{\Be\BM\BWi\Ahi\GKRRTZ\Ahii{--}\GP}.

We are interested in also the geometry of supermanifolds \refs{\RW,\Zho}. 
Topological sigma models are useful tools to study geometric properties of the 
manifolds. In constructing topological B-models, the Calabi-Yau condition is necessary.
Though ordinary complex projective space $\Bbb{CP}^3$ 
is not a Calabi-Yau manifold, the super twistor space $\Bbb{CP}^{3|4}$ 
is a super Calabi-Yau manifold \Se\ 
and topological B-model can be realised on this supermanifold $\Bbb{CP}^{3|4}$.
As a concrete example, 
a twistorial Calabi-Yau supermanifold was considered 
by \refs{\NV,\NOV}, in which S-duality of topological string was discussed. 

Topological sigma models have rich structures and 
it has been known that 
there is mirror symmetry of Calabi-Yau manifolds, which is 
the correspondence between the topological A-model
and the B-model (for example, see \refs{\HV,\HIV} and references therein).
The mirror symmetry for the supermanifolds was an interesting subject
and it was discussed in \refs{\Se,\AV\KP\Ah{--}\BDRSS}.

The gauged linear sigma model on an ordinary space was studied in \Wii\ and 
phases of the model were investigated. It has been believed there is 
correspondence between a Calabi-Yau and a Landau-Ginzburg theory, 
which is interpreted  as the phases of this gauged linear sigma model.
If a Fayet-Iliopoulos (FI) parameter $r$ becomes sufficiently large,  
the linear sigma model is reduced to 
the nonlinear sigma model on a Calabi-Yau manifold. 
On the other hand, for $r \ll 0$, the linear sigma model is reduced 
to a Landau-Ginzburg theory. 
This correspondence is understood as a kind of mirror symmetry. 
We expect that such correspondence governed by the FI parameter will be  
found in the gauged linear sigma model on a supermanifold.

D-branes in the topological string theory were considered in \OOY\ and 
there exist A-branes (B-branes) in the topological A-model (B-model). 
Since the D-branes are provided by the boundaries of open strings, 
the D-branes in the topological theory were derived from the analysis 
of boundary conditions preserving some supersymmetry. 
The gauged linear sigma model is useful tool to understand  
such D-branes \refs{\GJS,\GJSi}.
We are now interested in the D-branes embedded in the supermanifold, 
which will be derived from the supersymmetric boundary conditions in 
a gauged linear sigma model on the supermanifold.

In this paper, we shall consider a gauged linear sigma model on $\Bbb{CP}^{m-1|n}$. 
In section 2, 
we construct the Lagrangian of the $U(1)$ gauged linear sigma model on this manifold.
Here fermionic chiral superfields are introduced, 
whose lowest components are Grassmann odd fields. 
In section 3, the structure of the classical vacua of this 
model is analysed.  
Then we compute the one-loop correction to the D-field in order to study quantum effects
and discuss relations to the Ricci flatness condition for $\Bbb{CP}^{m-1|n}$. 
In section 4, we concentrate on open string sectors.
Here we discuss the A-type and the B-type boundary conditions and 
show the existence of A-branes and B-branes.
They are realisations of supersymmetric boundary conditions. 
In section 5, we consider two extensions of the $U(1)$ gauged linear sigma model. 
One is the non-abelian gauged linear sigma model and the other is 
the model with $(0,2)$ supersymmetry.
Section 6 is devoted to conclusions.

\newsec {$U(1)$ gauged linear sigma model}

We want to construct a gauged linear sigma model on a supermanifold.
Here we consider a $U(1)$ gauge group for simplicity.
Let us begin writing down the notations following 
\refs{\Wii,\WB}.
The supercoordinates of the world-sheet are 
denoted by $(x_0,x_1,\theta^\pm,\Btheta^\pm)$. 
We set the metric of the world-sheet to be $\eta_{ij} = {\rm diag}(-1,+1)$.
The differential operators are 
\eqn\Sder{\eqalign{
D_+ = {\partial \over \partial\theta^+} - i\Btheta^+(\partial_0+\partial_1), &\quad
D_- = {\partial \over \partial\theta^-} - i\Btheta^-(\partial_0-\partial_1), \cr
\BD_+ = -{\partial \over \partial\Btheta^+} + i\theta^+(\partial_0+\partial_1), &\quad
\BD_- = -{\partial \over \partial\Btheta^-} + i\theta^-(\partial_0-\partial_1),
}}
and the super charges are defined:
\eqn\Scharge{\eqalign{
Q_+ = {\partial \over \partial\theta^+} + i\Btheta^+(\partial_0+\partial_1), &\quad
Q_- = {\partial \over \partial\theta^-} + i\Btheta^-(\partial_0-\partial_1), \cr
\BQ_+ = -{\partial \over \partial\Btheta^+} - i\theta^+(\partial_0+\partial_1), &\quad
\BQ_- = -{\partial \over \partial\Btheta^-} - i\theta^-(\partial_0-\partial_1).
}}

We introduce two types of chiral superfields, namely, 
a bosonic chiral superfield and a new fermionic chiral superfield \Wi. 
Firstly an ordinary bosonic chiral superfield 
is defined by $\BD \Phi = 0$ and is described in terms of component fields
as 
\eqn\bcs{\eqalign{
\Phi &= \phi + \sqrt{2}(\theta^+\psi_+ + \theta^-\psi_-)
+ 2\theta^+\theta^- F \cr
&\phantom{=} -i\theta^-\Btheta^-(\partial_0-\partial_1)\phi 
-i\theta^+\Btheta^+(\partial_0+\partial_1)\phi 
- \theta^+\theta^-\Btheta^-\Btheta^+ (\partial_0^2 - \partial_1^2)\phi \cr
&\phantom{=} -i\sqrt{2}\theta^+\theta^-\Btheta^-(\partial_0-\partial_1)\psi_+
+i\sqrt{2}\theta^+\theta^-\Btheta^+(\partial_0+\partial_1)\psi_- .
}}
Here $\phi$ and $F$ are bosons, while $\psi_+$ and $\psi_-$ are fermions. 
Secondly we can define an anti-chiral superfield 
by $D\BPhi=0$. 
It is hermitian conjugate of the chiral superfield 
and is written as
\eqn\bacs{\eqalign{
\BPhi &= \Bphi - \sqrt{2}(\Btheta^+\Bpsi_+ + \Btheta^-\Bpsi_-)
+ 2\Btheta^-\Btheta^+ \BF \cr
&\phantom{=} +i\theta^-\Btheta^-(\partial_0-\partial_1)\Bphi 
+i\theta^+\Btheta^+(\partial_0+\partial_1)\Bphi 
- \theta^+\theta^-\Btheta^-\Btheta^+ (\partial_0^2 - \partial_1^2)\Bphi \cr
&\phantom{=} -i\sqrt{2}\theta^-\Btheta^-\Btheta^+(\partial_0-\partial_1)\Bpsi_+
+i\sqrt{2}\theta^+\Btheta^-\Btheta^+(\partial_0+\partial_1)\Bpsi_- .
}}
In the following sections, the lowest components 
$\phi$'s ($\Bphi$'s) of chiral (anti-chiral) superfields
are interpreted as bosonic coordinates of the target space. 
In the case of the supermanifold, 
we should introduce fermionic chiral superfields whose lowest components 
are Grassmann odd fields, namely, 
these lowest components describe the anti-commutative coordinates of the supermanifold.

In the same way, we can define the fermionic chiral superfield 
in component expansion: 
\eqn\fcs{\eqalign{
\Xi &= \xi + \sqrt{2}(\theta^+ b_+ + \theta^- b_-)
+ 2\theta^+\theta^- \chi \cr
&\phantom{=} -i\theta^-\Btheta^-(\partial_0-\partial_1)\xi 
-i\theta^+\Btheta^+(\partial_0+\partial_1)\xi 
- \theta^+\theta^-\Btheta^-\Btheta^+ (\partial_0^2 - \partial_1^2)\xi \cr
&\phantom{=} -i\sqrt{2}\theta^+\theta^-\Btheta^-(\partial_0-\partial_1)b_+
+i\sqrt{2}\theta^+\theta^-\Btheta^+(\partial_0+\partial_1)b_- .
}}
$\xi$ and $\chi$ are Grassmann odd, while $b_+$ and $b_-$ are Grassmann even. 
So this chiral superfield behaves totally as a Grassmann odd superfield.
The anti-chiral superfield is then represented by 
\eqn\facs{\eqalign{
\BXi &= \Bxi + \sqrt{2}(\Btheta^+ \Bb_+ + \Btheta^- \Bb_-)
+ 2\Btheta^-\Btheta^+ \Bchi \cr
&\phantom{=} +i\theta^-\Btheta^-(\partial_0-\partial_1)\Bxi 
+i\theta^+\Btheta^+(\partial_0+\partial_1)\Bxi 
- \theta^+\theta^-\Btheta^-\Btheta^+ (\partial_0^2 - \partial_1^2)\Bxi \cr
&\phantom{=} +i\sqrt{2}\theta^-\Btheta^-\Btheta^+(\partial_0-\partial_1)\Bb_+
-i\sqrt{2}\theta^+\Btheta^-\Btheta^+(\partial_0+\partial_1)\Bb_- .
}}
Note that the signs in front of some components in the fermionic 
anti-chiral superfield \facs\ are opposite to  
the corresponding components in the bosonic one \bacs. 
For example, compare the second term $-\sqrt{2}\,\Btheta^+\Bpsi_+$ 
on the right hand side in \bacs\ with $+\sqrt{2}\,\Btheta^+\Bb_+$ in \facs. 
The conjugate of $\theta^+\psi_+$ is $-\Btheta^+\Bpsi_+$ because 
$\Btheta$ and $\Bpsi$ are anti-commutative. On the other hand, 
since $\Bb_+$ is a bosonic component, $\Btheta^+$ and $\Bb_+$
can commute with each other and 
the conjugate of $\theta^+b_+$ becomes $+\Btheta^+\Bb_+$. 

We shall take  
$m$ bosonic and $n$ fermionic coordinates and consider 
an $(m|n)$-dimensional supermanifold.
Since these coordinates are regarded as the lowest components 
of chiral superfields, we introduce $m$ bosonic chiral superfields and 
$n$ fermionic chiral superfields,
$$
\Phi^I(x),\quad (I=1,2,\cdots,m),\qquad \Xi^A(x),\quad (A=1,2,\cdots,n).
$$

To construct a projective supermanifold, we consider the $U(1)$ gauge theory 
by introducing an abelian vector superfield $V$.
By the use of component fields, 
the vector superfield is described in the Wess-Zumino gauge,
\eqn\vects{\eqalign{
V &= -\sqrt{2}\theta^-\Btheta^+\sigma - \sqrt{2}\theta^+\Btheta^-\Bsigma
+ \theta^-\Btheta^-(v_0-v_1) + \theta^+\Btheta^+(v_0+v_1) \cr
&\phantom{=} - 2i\theta^+\theta^-(\Btheta^+\Blambda_+ + \Btheta^-\Blambda_-)
-2i\Btheta^-\Btheta^+(\theta^+\lambda_++\theta^-\lambda_-)
+ 2\theta^+\theta^-\Btheta^-\Btheta^+D .
}}
This superfield contains the auxiliary field $D$ which plays an important role 
in discussing vacua of the theory.
We then assign $U(1)$ charges $Q_I$ to $\Phi^I$ and $q_A$ to $\Xi^A$.

Now let us write down the Lagrangian of 
${\cal N} = (2,2)$ gauged linear sigma model on the $(m-1|n)$-dimensional 
target space.
The Lagrangian consists of four parts, which are
the kinetic part of chiral superfields $L_{\rm kin}$,
the kinetic part of gauge fields $L_{\rm gauge}$, 
the FI term with the theta parameter $L_{D,\theta}$  
and the superpotential term $L_W$. 
Then the total Lagrangian is the sum of these parts,
\eqn\Lag{
L = L_{\rm kin} + L_{\rm gauge} + L_{D,\theta} + L_W .
}

Firstly we study the kinetic part of the chiral superfields. 
The Lagrangian for these chiral superfields consists of two parts given by 
\eqn\Lagkin{
L_{\rm kin} = L_{\rm kin}^b + L_{\rm kin}^f, 
}
where $L_{\rm kin}^b$ and $L_{\rm kin}^f$ are the Lagrangians 
for the bosonic chiral superfields $\Phi^I$ and for the 
fermionic ones $\Xi^A$ respectively. 
In the gauged linear sigma model in two dimensions, 
the kinetic part of the Lagrangian of bosonic chiral superfields $L_{\rm kin}^b$
is well known \Wii. Calculating 
$L_{\rm kin}^b = \int d^2\theta d^2\Btheta \sum_{I=1}^m\BPhi^Ie^{2Q_IV}\Phi^I
$
by the use of \bcs, \bacs\ and \vects, we obtain the expression in terms of  component fields
\eqn\Lkinb{\eqalign{L_{\rm kin}^b 
&= \sum_I \biggl[ \BD_0\Bphi^I D_0\phi^I - \BD_1\Bphi^I D_1\phi^I
+ i \Bpsi^I_-(D_0+D_1)\psi^I_- + i \Bpsi^I_+(D_0-D_1)\psi^I_+ \cr
&\phantom{=}+ \BF^I F^I -\sqrt{2}Q_I(\sigma\Bpsi^I_-\psi^I_+ 
+ \Bsigma\Bpsi^I_+\psi^I_-)
-i\sqrt{2}Q_I\Bphi^I(\psi^I_-\lambda_+ - \psi^I_+\lambda_-) \cr
&\phantom{=}-i\sqrt{2}Q_I\phi^I(\Blambda_-\Bpsi^I_+ - \Blambda_+\Bpsi^I_-) 
+ (Q_ID - 2Q_I^2\Bsigma\sigma)\Bphi^I\phi^I
\biggr] ,
}}
where $D_j$ and $\BD_j$ $(j=0,1)$ are covariant derivatives described as  
$D_j = \partial_j + iQ_Iv_j$ and $\BD_j = \partial_j - iQ_Iv_j$.

In the same way, we can define the Lagrangian for the fermionic chiral 
superfields $\Xi^A$:
$$
L_{\rm kin}^f = \int d^2\theta d^2\Btheta \sum_{A=1}^n\BXi^Ae^{2q_AV}\Xi^A. 
$$
Substituting \fcs, \facs\ and \vects\ into this equation, 
the Lagrangian for $\Xi^A$ becomes 
\eqn\Lkinf{\eqalign{L_{\rm kin}^f 
&= \sum_A \biggl[ \BD_0\Bxi^A D_0\xi^A - \BD_1\Bxi^A D_1\xi^A 
	+ i \Bb^A_- (D_0+D_1)b^A_- + i \Bb^A_+(D_0-D_1)b^A_+ \cr
&\phantom{=}+ \Bchi^A\chi^A -\sqrt{2}q_A(\sigma\Bb^A_-b^A_+ 
	+ \Bsigma\Bb^A_+b^A_-)
	+i\sqrt{2}q_A\Bxi^A(b^A_-\lambda_+-b^A_+\lambda_-) \cr
&\phantom{=}-i\sqrt{2}q_A\xi^A(\Blambda_-\Bb^A_+ - \Blambda_+\Bb^A_-) 
	+ (q_AD - 2q_A^2\Bsigma\sigma)\Bxi^A\xi^A
\biggr] ,
}}
with  
$D_j = \partial_j + iq_Av_j$ and $\BD_j = \partial_j - iq_Av_j$.
We should note that the spin of $\xi$ is equal to zero 
though $\xi$ is a Grassmann odd field.  
In this paper, the word  ``{\it fermionic}" often means 
merely {\it Grassmann odd}.

Secondly we write down the kinetic part of the gauge fields \Wii. 
For this purpose, it is useful to define field strength $\Sigma$ 
of the vector superfield as $\Sigma = (1 / \sqrt{2})\BD_+D_-V$. 
In component 
expansion, it is represented as
$$\eqalign{
\Sigma &= \sigma + i\sqrt{2}\theta^+\Blambda_+ - i\sqrt{2}\Btheta^-\lambda_-
+ \sqrt{2}\theta^+\Btheta^- (D - iv_{01}) 
+ i\theta^-\Btheta^-(\partial_0-\partial_1)\sigma \cr
&\phantom{=}- i\theta^+\Btheta^+(\partial_0+\partial_1)\sigma 
- \sqrt{2}\theta^+\theta^-\Btheta^-(\partial_0-\partial_1)\Blambda_+
+ \sqrt{2}\theta^+\Btheta^-\Btheta^+(\partial_0+\partial_1)\lambda_- \cr
&\phantom{=}+\theta^+\theta^-\Btheta^-\Btheta^+
(\partial_0^2-\partial_1^2)\sigma ,
}$$
with $v_{01} = \partial_0v_1 - \partial_1v_0$. 
This field strength is the twisted chiral superfield and 
the kinetic part of the gauge fields is denoted by
\eqnn\Lgauge
$$\eqalignno{
L_{\rm gauge} &= - {1 \over e^2}\int d^4\theta\ \BSigma\Sigma \cr
	&= {1 \over e^2} \biggl( {1 \over 2}{v_{01}}^2 
		+ {1 \over 2}D^2 + i\Blambda_+(\partial_0-\partial_1)\lambda_+
		+ i\Blambda_-(\partial_0+\partial_1)\lambda_-
		-\partial_i\Bsigma\partial^i\sigma \biggr) , &\Lgauge
}$$
where $e$ is a gauge coupling constant.

Thirdly we can introduce the FI term with 
the theta parameter in the supersymmetric theory \Wii. 
In terms of $t = ir + \theta /(2\pi)$, 
the Lagrangian $L_{D,\theta}$ becomes 
\eqnn\LDtheta
$$\eqalignno{
L_{D,\theta} &= 
{it\over 2\sqrt{2}}\int d\theta^+d\Btheta^-\Sigma \big|_{\theta^-=\Btheta^+=0}
- {i{\bar t}\over 2\sqrt{2}}\int d\theta^-d\Btheta^+\BSigma \big|_{\theta^+=\Btheta^-=0} \cr
&= -rD + {\theta \over 2\pi} v_{01} . &\LDtheta
}$$
$r$ is called the FI parameter. 
Note that the theta angle $\theta$ has periodicity of $2\pi$ in physics \Wii.

Next we turn to  the superpotential terms constructed by the use of 
holomorphic polynomials $W(\Phi,\Xi)$ of the chiral superfields.
The Lagrangian $L_W$ is 
described as 
\eqn\LWdef{
L_W = - {1 \over 2} \int d\theta^-d\theta^+ 
W(\Phi,\Xi)\big|_{\Btheta^+=\Btheta^-=0} 
- {1 \over 2} \int d\Btheta^+d\Btheta^- 
\BW(\BPhi,\BXi)\big|_{\theta^+=\theta^-=0} .
}
When the superpotential includes only bosonic chiral superfields, 
derivatives with respect to only $\phi$ and $\Bphi$ appear \refs{\Wii,\WB}.
On the other hand, 
we have to pay attention to derivatives with respect to Grassmann odd variables.
When we exchange  Grassmann odd variables one another, there appear extra signs. 
From now on, 
we use only left derivatives for Grassmann odd variables. 
For example, if the superpotential is a monomial of $\Xi^A$ 
with degree $p$, $W(\Phi,\Xi) = \Xi^{A_1}\Xi^{A_2}\cdots \Xi^{A_p}$, 
then the Lagrangian is calculated from \LWdef\ 
$$\eqalign{
L_W &= - {1 \over 2} \int d\theta^-d\theta^+ \Biggl(
	\sum_{i=1}^p (-1)^{i-1}2\theta^+\theta^-
	\chi^{A_i}\xi^{A_1}\cdots\xi^{A_{i-1}}\xi^{A_{i+1}}\cdots\xi^{A_p} \cr
&\phantom{=} + \sum_{i<j}(-1)^{i+j-1}
	2\theta^+\theta^-(b_+^{A_i}b_-^{A_j} - b_-^{A_i}b_+^{A_j})
	\xi^{A_1}\cdots\xi^{A_{i-1}}\xi^{A_{i+1}}\cdots
	\xi^{A_{j-1}}\xi^{A_{j+1}}\cdots\xi^{A_p} \Biggr) \cr
&\phantom{=}-{1 \over 2}\int d\Btheta^+d\Btheta^-\biggl(
	\sum_{i=1}^p (-1)^{i-1}2 
	\Bxi^{A_p}\cdots\Bxi^{A_{i+1}}\Bxi^{A_{i-1}}\cdots\Bxi^{A_1}
	\Bchi^{A_i}\Btheta^-\Btheta^+ \cr
&\phantom{=}+ \sum_{i<j}(-1)^{i+j-1}
	2\Bxi^{A_p}\cdots\Bxi^{A_{j+1}}\Bxi^{A_{j-1}}\cdots\Bxi^{A_{i+1}}\Bxi^{A_{i-1}}\cdots\Bxi^{A_1} 
	\Btheta^-\Btheta^+(\Bb_+^{A_i}\Bb_-^{A_j} - \Bb_-^{A_i}\Bb_+^{A_j}) \Biggr) \cr
&= - \Biggl(
	\sum_{i=1}^p\chi^{A_i}{\partial \over \partial \xi^{A_i}}(\xi^{A_1}\cdots\xi^{A_p})
	+ \sum_{i,j=1}^p b_-^{A_i}b_+^{A_j}
	{\partial \over \partial \xi^{A_i}}{\partial \over \partial \xi^{A_j}}
	(\xi^{A_1}\cdots\xi^{A_p}) \Biggr) \cr
&\phantom{=} - \Biggl(
	\sum_{i=1}^p\Bchi^{A_i}{\partial \over \partial \Bxi^{A_i}}
	(\Bxi^{A_p}\cdots\Bxi^{A_1})
	+ \sum_{i,j=1}^p \Bb_+^{A_i}\Bb_-^{A_j}
	{\partial \over \partial \Bxi^{A_i}}{\partial \over \partial \Bxi^{A_j}}
	(\Bxi^{A_p}\cdots\Bxi^{A_1}) \Biggr) .
}$$ 
If we use only left derivatives for  
the superpotential of fermionic chiral superfields, 
we can  apply the same notation as the bosonic case \Wii. 
In the same way, we can evaluate \LWdef\ for the general superpotential 
\eqn\LW{\eqalign{
L_W &= - \biggl[ \biggl(
	\sum_I F^I{\partial W(\phi,\xi) \over \partial \phi^I} 
	+ \sum_{I,J} \psi_-^I\psi_+^J
	{\partial^2 W(\phi,\xi) \over \partial \phi^I\partial\phi^J} \cr
&\phantom{=}+ \sum_A \chi^A {\partial W(\phi,\xi) \over \partial \xi^A} 
+ \sum_{A,B} b_-^Ab_+^B {\partial^2 W(\phi,\xi) \over \partial \xi^A\partial\xi^B} 
	- \sum_{I,A} (\psi_+^Ib_-^A - \psi_-^Ib_+^A)
	{\partial^2 W(\phi,\xi) \over \partial \phi^I\partial \xi^A} \biggr) \cr
&\phantom{=}+ \biggl( \sum_I \BF^I
	{\partial \BW(\Bphi,\Bxi) \over \partial \Bphi^I} 
	+ \sum_{I,J} \Bpsi_+^I\Bpsi_-^J
	{\partial^2 \BW(\Bphi,\Bxi) \over \partial \Bphi^I\partial\Bphi^J} \cr
&\phantom{=}+ \sum_A \Bchi^A {\partial \BW(\Bphi,\Bxi) \over \partial \Bxi^A} 
	+ \sum_{A,B} \Bb_+^A\Bb_-^B 
	{\partial^2 \BW(\Bphi,\Bxi) \over \partial \Bxi^A\partial\Bxi^B} 
	- \sum_{I,A} (\Bpsi_+^I\Bb_-^A - \Bpsi_-^I\Bb_+^A)
	{\partial^2 \BW(\Bphi,\Bxi) \over \partial \Bphi^I\partial \Bxi^A} \biggr)
\biggr] .
}}
This Lagrangian provides the interaction terms 
between the Grassmann even field $\phi$ and the Grassmann odd one $\xi$. From the viewpoint 
of the target space geometry, such terms are associated 
with a subsupermanifold embedded in the target supermanifold.

\newsec{Phases of classical vacuum and one-loop correction}

The structure of the classical vacua of 
the gauged linear sigma model on the ordinary bosonic manifold was 
considered in \Wii.
This model has two phases. One phase is the sigma model on 
Calabi-Yau hypersurfaces in complex projective space 
for the FI parameter $r \gg 0 $. 
The other is the Landau-Ginzburg theory for $r \ll 0$.

In this section, we study the structure of the classical vacua of 
the $U(1)$ gauged linear sigma model on the supermanifold.
In discussing the vacua, 
the potential energy of the dynamical fields $\phi$ and $\xi$ is important. 
We can see this potential from the Lagrangian \Lag\ and it is described as
\eqn\potI{
U = \sum_I\BF^IF^I + \sum_A \Bchi^A\chi^A 
+ 2|\sigma|^2\biggl( \sum_I Q_I^2 |\phi^I|^2 + \sum_A q_A^2 \Bxi^A\xi^A \biggr)
+ {1 \over 2e^2}D^2.
}
Since $F$, $\chi$ and $D$ are auxiliary fields, we can integrate out them 
in terms of the equations of motion:
\eqna\eomaux
$$\eqalignno{
&D = -e^2 \biggl( \sum_IQ_I|\phi^I|^2 + \sum_Aq_A\Bxi^A\xi^A - r \biggr), 
&\eomaux a\cr
&\BF^I = {\partial W \over \partial \phi^I}, \quad 
F^I = {\partial \BW \over \partial \Bphi^I}, &\eomaux b \cr
&\Bchi^A = - {\partial W \over \partial \xi^A}, \quad
\chi^A = {\partial \BW \over \partial \Bxi^A}. &\eomaux c
}$$
 \eomaux{b} implies 
that the superpotential $W$ should be Grassmann even 
because $\BF$ and $\phi$ are Grassmann even.
This condition is consistent with \eomaux{c};
if the superpotential is Grassmann odd, the sigma model might be ill-defined.
Substituting \eomaux{} into \potI, 
the potential energy including only $\phi$, $\xi$ and $\sigma$ 
is written as
\eqn\potII{\eqalign{
U &= \sum_I{\partial \BW \over \partial \Bphi^I}{\partial W \over \partial \phi^I}
+ \sum_A {\partial \BW \over \partial \Bxi^A}{\partial W \over \partial \xi^A}
+ 2|\sigma|^2\biggl( \sum_I Q_I^2 |\phi^I|^2 + \sum_A q_A^2 \Bxi^A\xi^A \biggr) \cr
&\phantom{=}+ {e^2 \over 2}\biggl(\sum_IQ_I|\phi^I|^2 + \sum_Aq_A\Bxi^A\xi^A - r \biggr)^2 .
}}
Let us now analyse the structure of the classical vacua $U=0$.

As a first example, we shall consider the case without superpotential. 
In such a case, the potential $U$ is evaluated as
$$
U = 2 |\sigma|^2\biggl(
	\sum_{I=1}^mQ_I^2|\phi^I|^2+
\sum_{A=1}^nq_A^2 \Bxi^A\xi^A \biggr) 
+ {e^2\over 2}\biggl(\sum_{I=1}^mQ_I|\phi^I|^2+\sum_{A=1}^n
q_A \Bxi^A \xi^A -r\biggr)^2 .
$$
In order for this potential to vanish, we have to impose $\sigma = 0$ and 
\eqn\Dconst{
\biggl(-{D \over e^2} =\biggr) 
\sum_I Q_I |\phi^I|^2 + \sum_Aq_A\Bxi^A\xi^A - r = 0 .
}
This equation is usually called the D-flatness condition.
For $r\gg 0$, there should be positive charges in $Q_I$ or $q_A$. 
For simplicity we set the model with the positive charges $Q_I, q_A>0$ 
for all $I$ and $A$. 
Then the condition \Dconst\ for vacua represents 
the weighted projective space 
$\Bbb{WCP}^{m-1|n}_{(Q_1,Q_2,\cdots,Q_m|q_1,q_2,\cdots,q_n)}$.

Next we shall consider another example.
We introduce a neutral chiral superfield $P$ with $Q_P=0$ and 
set the superpotential 
$W=P\cdot G(\Phi,\Xi)$. 
Here $G$ is a transversal homogeneous polynomial in the same way as \Wii. 
Then the potential $U$ is written as
\eqn\expotII{\eqalign{
U &=  |p|^2 \biggl(
\sum_{I=1}^m
{\partial \BG \over \partial \Bphi^I}
{\partial G\over \partial \phi^I}
+\sum_{A=1}^n
{\partial \BG \over \partial \Bxi^A}
{\partial G\over \partial \xi^A}
\biggr)  + |G|^2 \cr
&\phantom{=} + 2|\sigma|^2
\biggl(\sum_{I=1}^mQ_I^2|\phi^I|^2+
\sum_{A=1}^nq_A^2 \Bxi^A\xi^A\biggr) 
+ {e^2 \over 2}\biggl(\sum_{I=1}^m Q_I|\phi^I|^2 
+ \sum_{A=1}^n q_A \Bxi^A\xi^A - r \biggr)^2 . 
}}
In order for this potential to vanish, we have to impose 
the D-flatness condition \Dconst. 
By the same analysis in the previous example, for $r\gg 0$ there should 
be positive charges in $Q_I$ or $q_A$ and 
we then take the positive charges $Q_I, q_A>0$ for all $I$ and $A$. 
If $p \neq 0$, $\partial G / \partial \phi^I$ and 
$\partial G / \partial \xi^A$ must vanish simultaneously. 
Then all $\phi^I$ and $\xi^A$ become zero
because of the transversality of the superpotential and 
the D-flatness condition is not satisfied. So the conditions for 
the potential energy \expotII\ to vanish are realised as 
$$
\sum_IQ_I|\phi^I|^2 + \sum_Aq_A\Bxi^A\xi^A - r = 0,\quad 
\sigma =0,\quad p=0,\quad G=0 .
$$
They represent the hypersurface $G=0$ in $\Bbb{WCP}^{m-1|n}$.

Another intriguing example is 
the model in which the superpotential $W(\Phi,\Xi)$ is 
factorized as $W(\Phi,\Xi) = W_b(\Phi)W_f(\Xi)$.
Here $W_b(\Phi)$ and $W_f(\Xi)$ are respectively  
transversal homogeneous polynomials of $\Phi$ and $\Xi$. 
In order for the last term in \potII\ to vanish, 
we should take the D-flatness condition \Dconst. 

Firstly we consider the case $r \gg 0$. 
Since there should be positive charges in $Q_I$ or $q_A$ 
in order for \Dconst\ to be satisfied, we consider the case that 
all the charges $Q_I$ and $q_A$ are positive. From \potII\ and \Dconst, 
the conditions for the potential energy are described as 
\eqn\Wconstp{
\sigma = 0,\quad
W_f(\xi){\partial W_b(\phi) \over \partial \phi^I}=0,\quad
W_b(\phi){\partial W_f(\xi) \over \partial \xi^A}=0.
}
If $\partial W_b / \partial \phi^I$ and 
$\partial W_f / \partial \xi^A$ vanish simultaneously, 
all $\phi^I$ and $\xi^A$ become zero
on account of transversality of the superpotential. 
Then the D-flatness condition \Dconst\ is not satisfied for $r \gg 0$.
Hence the solutions of the constraints \Wconstp\ is classified into the 
following three cases:

\item{I.}$W_b=0,\quad W_f=0$ \hfill\break
Since there are the $U(1)$ gauge symmetry,
$\phi^I \to e^{iQ_I\alpha}\phi^I$ and $\xi^A \to e^{iq_A\alpha}\xi^A$,
and the symmetry under the scaling by $r$ in \Dconst, we can take the identification denoted by
$$
(\phi^1,\phi^2,\cdots,\phi^m|\xi^1,\xi^2,\cdots,\xi^n) \sim 
(\lambda^{Q_1}\phi^1,\lambda^{Q_2}\phi^2,\cdots,\lambda^{Q_m}\phi^m|
\lambda^{q_1}\xi^1,\lambda^{q_2}\xi^2,\cdots,\lambda^{q_n}\xi^n), 
$$
where $\lambda \in \Bbb{C}^\times$.
In other words, the vacuum manifold becomes 
the super weighted complex projective space 
$\Bbb{WCP}^{m-1|n}_{(Q_1,Q_2,\cdots,Q_m|q_1,q_2,\cdots,q_n)}$.
The theory is reduced to the sigma model on the surface given by the 
equations $W_b=0$ and $W_f=0$ in this supermanifold 
$\Bbb{WCP}^{m-1|n}_{(Q_1,Q_2,\cdots,Q_m|q_1,q_2,\cdots,q_n)}$. 
\item{II.}${\partial W_f \over \partial \xi^A} = 0, \quad W_f= 0$ \hfill\break
On account of the transversality of the superpotential $W_f$, 
the equation $\partial W_f/\partial \xi^A = 0$ leads to 
$\xi^A=0$ for all $A$. 
On the other hand, the bosonic fields $\phi^I$'s live 
on the surface defined by $\sum_IQ_I\Bphi^I\phi^I=r$. 
The $U(1)$ gauge symmetry is then completely broken in general. 
But the vacua may have residual gauge symmetry 
in specific cases. For example,  if all the charges of the fermionic 
fields $\xi^A$'s have the same value, that is, 
$q_1=q_2=\cdots=q_m \equiv q$, 
there exists the residual $\Bbb{Z}_q$ symmetry described as
$\phi^I \to e^{2\pi i Q_I/q}\phi^I$.
As a result, this theory is reduced to the ordinary (bosonic) 
$\Bbb{Z}_q$ orbifold. This phase is just an analogue of 
the Landau-Ginzburg orbifold derived from the ordinary $\Bbb{CP}^{m-1}$ model in \Wii.
\item{III.}${\partial W_b \over \partial \phi^I} = 0, \quad W_b= 0$ \hfill\break
In this sector, a new feature appears.
By the transversality of the superpotential $W_b$, 
the only solution of $\partial W_b / \partial \phi^I=0$ 
is $\phi^I=0$ for all $I$ and \Dconst\ becomes $\sum_Aq_A\Bxi^A\xi^A=r$.
It means the sigma model leads to the theory on the quadratic surface 
given by $\sum_Aq_A\Bxi^A\xi^A=r$ in the purely fermionic manifold.
When we set $Q_1=Q_2=\cdots=Q_m\equiv Q$, the original $U(1)$ symmetry is 
reduced to $\Bbb{Z}_{Q}$ symmetry 
in the same way as the case II and the target space 
can be regarded as $\Bbb{Z}_{Q}$ orbifold.

\noindent
From the vacuum structures in the cases II and III, 
there might be a transition between 
the models on the bosonic and the fermionic manifolds. 

Next we consider the case $r \ll 0$. 
Since some charges in $Q_I$ and $q_A$ should be negative from \Dconst, 
we assume that all charges are negative, $Q_I, q_A<0$, for simplicity.
We can then find the solutions for $U=0$ expressed by
the equations \Wconstp. By the same analysis as the $r \gg 0$ case, 
there are three kinds of solutions for \Wconstp:
the theory on the surface $W_b=0=W_f$ in the super
weighted projective space 
$\Bbb{WCP}^{m-1|n}_{(Q_1,Q_2,\cdots,Q_m|q_1,q_2,\cdots,q_n)}$, 
the theory on the quadratic surface $\sum_IQ_I\Bphi^I\phi^I=r$ in the 
ordinary Grassmann even manifold and 
the theory on the quadratic surface $\sum_Aq_A\Bxi^A\xi^A=r$ in the 
manifold with only the Grassmann odd coordinates.

In the above analyses, the flip of the sign of FI parameter $r$ is 
synonymous with the flip of the sign of all $U(1)$ charges.
At any rate, in response to the conditions for the superpotential, 
there are the three kinds of classical vacua; 
the supermanifold $\Bbb{WCP}$, 
the pure bosonic manifold and the pure fermionic manifold for this simple example.

\subsec{one-loop correction}
The D-flatness condition \Dconst\ has played an important role in studying classical vacua in 
the previous discussion. 
The sigma model receives quantum corrections and 
we should analyse quantum properties of the associated vacua.
Now we study the one-loop correction to the D-field.
The expectation value of the D-field diverges on account of 
the correction induced by the one-loop diagrams of 
the fields $\phi$ and $\xi$. From the Lagrangian \Lkinb, 
we can calculate the one-loop correction by $\phi$ as 
\eqn\oneloopb{
\biggl\langle -{D \over e^2} \biggr\rangle_{\hbox{\sevenrm 1-loop}}^{\rm bos.}
= \sum_I Q_I \int {d^2k \over (2\pi)^2} {1 \over k^2 + 2Q_I^2\Bsigma\sigma}
= {1 \over 4\pi} \sum_I Q_I \log \biggl({\mu^2 \over 2Q_I^2\Bsigma\sigma}\biggr),
}
where $\mu$ is a cut-off parameter. 
For the purely bosonic manifold \refs{\Wii,\MP}\
,  the condition for
the one-loop divergence to vanish is $\sum_{I=1}^m Q_I=0$. 
For example, the model with the charges $(1,1,\cdots,1,-m+1)$
was considered in \refs{\Wii,\MP}.
On the other hand, the sum $\sum_{I=1}^mQ_I$ for 
the weighted projective space 
$\Bbb{WCP}^{m-1}_{(Q_1,Q_2,\cdots,Q_m)}$
is not vanishing and the associated model is not scale-invariant. 
Since the first Chern class of this manifold
is proportional to the sum $\sum_{I=1}^mQ_I$, 
the scale invariance means the Calabi-Yau condition for the manifold. 
For example, the 
projective space $\Bbb{CP}^{m-1}(=\Bbb{WCP}^{m-1}_{(1,1,\cdots,1)})$ is not a Calabi-Yau manifold.

Next let us see the Grassmann odd fields.
The one-loop correction  
of the Grassmann odd fields $\xi^A$ to the expectation value 
$\langle -D/e^2\rangle$ becomes
\eqn\oneloopf{
\biggl\langle -{D \over e^2} \biggr\rangle_{\hbox{\sevenrm 1-loop}}^{\rm fer.}
= -\sum_A q_A \int {d^2k \over (2\pi)^2} {1 \over k^2 + 2q_A^2\Bsigma\sigma}
= -{1 \over 4\pi} \sum_A q_A \log \biggl({\mu^2 \over 2q_A^2\Bsigma\sigma}\biggr).
}
Note that it is different from the one-loop result of the bosonic field $\phi$ because  
the Grassmann odd field $\xi$ leads to an extra $-1$ factor.

The total one-loop correction by the dynamical scalar fields $\phi$ and $\xi$ 
is expressed as the sum of \oneloopb\ and \oneloopf,
\eqn\oneloop{\eqalign{
\biggl\langle -{D \over e^2} \biggr\rangle_{\hbox{\sevenrm 1-loop}}
&= {1 \over 4\pi} \biggl( \sum_I Q_I - \sum_A q_A\biggr)\log \mu^2 \cr
&\phantom{=} -{1 \over 4\pi} \biggl[ \sum_I Q_I\log(2Q_I^2\Bsigma\sigma)
- \sum_A q_A \log(2q_A^2\Bsigma\sigma)\biggr] .
}}
From this equation, it turns out the one-loop divergence vanishes in the situation;
\eqn\CYcond{
\sum_{I=1}^mQ_I - \sum_{A=1}^n q_A = 0.
}
If all the charges are positive for the purely bosonic manifold, 
$\sum_IQ_I$ does not vanish. 
But, for the supermanifold, the cancellation between the bosonic and 
the fermionic sectors can be established 
even if all the charges $Q_I$ and $q_A$ 
are positive.

Here we shall consider an example.
In our sigma model, the $U(1)$ symmetry and the rescaling of $r$ provide 
the identification,
\eqn\CPid{
(\phi^1,\cdots,\phi^m|\xi^1,\cdots,\xi^n) \sim
(\lambda^{Q_1}\phi^1,\cdots,\lambda^{Q_m}\phi^m|
\lambda^{q_1}\xi^1,\cdots,\lambda^{q_n}\xi^n),\quad
\lambda \in \Bbb{C}^\times.
}
This leads the target space to the super weighted projective space
$\Bbb{WCP}^{m-1|n}_{(Q_1,\cdots,Q_m|q_1,\cdots,q_n)}$.
If $Q_I=q_A=1$ for all $I$ and $A$, this target space is reduced to the 
super complex projective space $\Bbb{CP}^{m-1|n}$.
The condition \CYcond\ then becomes $m-n=0$ and it is consistent with 
the Calabi-Yau condition for the supermanifold shown in \Se.
Namely $\Bbb{CP}^{m-1|m}$ is the Calabi-Yau supermanifold  
and in this case the one-loop correction \oneloop\ is equal to zero.
The Calabi-Yau condition is important  in constructing topological theories.
Recent progress on the twistor space was started with the topological 
B-model on Calabi-Yau $\Bbb{CP}^{3|4}$ in \Wi.

We return to the expression of the corrected D-field and 
show that the one-loop correction is absorbed by the FI parameter. 
We introduce the effective FI parameter $r_{\rm eff}$ as
\eqn\FIeff{
r_{\rm eff} = r - {1 \over 2\pi}
\biggl[\biggl(\sum_IQ_I - \sum_Aq_A\biggr)\log{\mu \over \sqrt{2}|\sigma|}
- \sum_IQ_I\log|Q_I| + \sum_Aq_A\log|q_A|\biggr] ,
}
and the one-loop corrected D-field from \eomaux{a}\ and 
\oneloop\ is then described as 
$$
\biggl\langle -{D \over e^2} \biggr\rangle 
= \sum_IQ_I|\phi^I|^2 + \sum_Aq_A\Bxi^A\xi^A - r_{\rm eff} .
$$
For the Calabi-Yau $\Bbb{CP}^{m-1|m}$, \FIeff\ means that $r_{\rm eff}$
is equal to the bare $r$ and is independent of $\sigma$. 
For the case of $\sum_IQ_I-\sum_Aq_A \neq 0$, 
$r_{\rm eff}$ absorbs the one-loop correction depending on the cut-off parameter 
and then becomes the function of $\sigma$.

Now let us study the relation to the dual theory proposed by 
\refs{\Se,\AV}.
The FI parameter and the theta parameter are combined into the parameter $t$, 
which appears in the twisted superpotential 
${\tilde W}(\Sigma)={it \over {2\sqrt{2}}}\Sigma$ 
with $t=ir+ \theta / (2\pi)$, 
$$
L_{D,\theta}={it\over {2\sqrt{2}}}\int d\theta^+ d\Btheta^- \Sigma + (\hbox{h.c.}) .
$$
By considering the modified $r_{\rm eff}$, 
we can obtain the corrected  
twisted superpotential 
${\tilde W}_{\rm eff}(\Sigma)$
$$
\tilde{W}_{\rm eff}(\Sigma)
={1\over {2\sqrt{2}}}\Sigma \biggl[
it-{1\over {2\pi}}\sum_IQ_I\log \biggl({{\sqrt{2}Q_I\Sigma}\over \mu} \biggr)
+{1\over {2\pi}}\sum_Aq_A\log \biggl({{\sqrt{2}q_A\Sigma}\over \mu} \biggr)
\biggr] .
$$
Here we evaluated 
the loop-correction of this potential, 
but there is a dual mirror theory \refs{\Se,\AV}.
When one introduces twisted (bosonic) chiral superfields 
$Y_I= \BPhi^I\Phi^I$, $Y_A=-\BXi^A\Xi^A$ 
and pairs of fermionic fields $\eta_A$ and $\chi_A$, 
the corresponding twisted superpotential is 
defined as
$$
\tilde{W} =\Sigma \biggl(\sum_IQ_IY_I-\sum_Aq_AY_A+it\biggr) 
+ {1\over {2\pi}}\sum_{I}{\mu\over {\sqrt{2}}}e^{-2\pi Y_I}
+{1\over {2\pi}}\sum_{A}{\mu\over {\sqrt{2}}}e^{-2\pi \tilde{Y}_A} , 
$$
where $\tilde{Y}_A=Y_A+\eta_A\chi_A$.
By integrating $Y_I$, $Y_A$, $\eta_A$ and $\chi_A$, 
the same effective potential ${\tilde W}_{\rm eff}$ is recovered. 

Next we shall take the extension of the gauge 
group to $U(1)^N =\otimes_{a=1}^N U(1)_a$ 
by introducing $N$ gauge fields $V_a$ 
(its field strength $\Sigma_a$) $(a=1,2,\cdots ,N)$ 
with gauge couplings $e_a$. 
Matter parts consist of bosonic chiral superfields 
$\Phi^{Ia}$ $(I=1,2,\cdots ,m)$ 
and fermionic chiral superfields $\Xi^{Aa}$ $(A=1,2,\cdots ,n)$. 
The associated $U(1)_a$ ($a=1,2,\cdots ,N$) charges of these fields are 
defined as $Q_{Ia}$ for $\Phi^{Ia}$ and $q_{Aa}$ for $\Xi^{Aa}$. 
The full action then can be written down as
$$\eqalign{
L &=L^b_{\rm kin}+L^f_{\rm kin}+L_{\rm gauge}+L_{D,\theta}+L_{W},\cr
&L_{\rm kin}^b = \int d^2\theta d^2\Btheta \sum_{a=1}^N\sum_{I=1}^m
\BPhi^{Ia}e^{2Q_{Ia}V_a}\Phi^{Ia},\quad 
L_{\rm kin}^f = \int d^2\theta d^2\Btheta \sum_{a=1}^N\sum_{A=1}^n\BXi^{Aa}e^{2q_{Aa}V_a}\Xi^{Aa},\quad \cr
&L_{\rm gauge} = - \sum_{a=1}^N{1 \over e^2_a}\int d^4\theta\ \BSigma_a\Sigma_a, \cr
&L_{D,\theta} = \sum_{a=1}^N{it_a\over 2\sqrt{2}}\int d\theta^+d\Btheta^-\Sigma_a \big|_{\theta^-=\Btheta^+=0}
- \sum_{a=1}^N{i{\bar t_a}\over 2\sqrt{2}}\int d\theta^-d\Btheta^+\BSigma_a \big|_{\theta^+=\Btheta^-=0}, \cr
&L_W = - {1 \over 2} \int d\theta^-d\theta^+ 
W(\Phi,\Xi)\big|_{\Btheta^+=\Btheta^-=0} 
- {1 \over 2} \int d\Btheta^+d\Btheta^- 
\BW(\BPhi,\BXi)\big|_{\theta^+=\theta^-=0} .
}$$
$t_a=ir_a+ \theta_a/ (2\pi)$ ($a=1,2,\cdots ,N$) is the FI parameter  
associated with $U(1)_a$ gauge symmetry. 
By doing the same analyses as the $U(1)$ case, 
we can obtain D-fields $D_a$'s and 
their one-loop corrections
$$\eqalign{
D_a &= -e^2_a \biggl(\sum_I Q_{Ia}\Bphi^{Ia}\phi^{Ia}
    +\sum_A q_{Aa}\Bxi^{Aa}\xi^{Aa}-r_a\biggr),
\quad a=1,2,\cdots ,N, \cr
\biggl\langle -{D_a \over e^2_a} \biggr\rangle_{\hbox{\sevenrm 1-loop}}
&= {1 \over 4\pi} \sum_I Q_{Ia} \log \biggl({\mu^2 \over 2Q_{Ia}^2\Bsigma_a\sigma_a}\biggr)
-{1 \over 4\pi} \sum_A q_{Aa} \log \biggl({\mu^2 \over 2q_{Aa}^2\Bsigma_a\sigma_a}\biggr).\cr
}$$
These lead us to corrected FI parameters 
$r_{{\rm eff},a}$ $(a=1,2,\cdots ,N)$, 
and induce twisted superpotential
${\tilde W}_{\rm eff}(\Sigma)$, 
$$
\tilde{W}_{\rm eff}(\Sigma)
=\sum_{a=1}^N{1\over 2\sqrt{2}}\Sigma_a  \biggl[ it_a
-{1\over {2\pi}}\sum_I Q_{Ia}\log \biggl({{\sqrt{2}Q_{Ia}\Sigma_a}\over \mu} \biggr)
+{1\over {2\pi}}\sum_Aq_{Aa}\log \biggl({{\sqrt{2}q_{Aa}\Sigma_a}\over \mu} \biggr)\biggr].
$$
In order to turn a dual mirror theory,
we introduce twisted (bosonic) chiral superfields 
$Y_{Ia}=\BPhi^{Ia}\Phi^{Ia}$, $Y_{Aa}=-\BXi^{Aa}\Xi^{Aa}$ 
and pairs of fermionic fields $\eta_{Aa}$ and $\chi_{Aa}$. 
The corresponding twisted superpotential in the mirror side is 
defined as
$$
{\tilde W} =\sum_{a=1}^N \Sigma_a 
\biggl(\sum_IQ_{Ia}Y_{Ia}-\sum_Aq_{Aa}Y_{Aa}+it_a\biggr) 
+ {1\over 2\pi}\sum_{I,a}{\mu \over \sqrt{2}}e^{-2\pi Y_{Ia}}
+ {1\over 2\pi}\sum_{A,a}{\mu \over \sqrt{2}}e^{-2\pi \tilde{Y}_{Aa}}, 
$$
where ${\tilde Y}_{Aa}=Y_{Aa}+\eta_{Aa}\chi_{Aa}$.

\newsec{Supersymmetric boundary condition}

D-branes are realised as the boundaries of open strings and preserve some supersymmetry. 
We can find the D-branes in the sigma model by considering 
the world-sheet with boundaries. 

We shall write down the supersymmetric transformations. 
By the use of the super charges \Scharge, 
the generator of the supersymmetric transformation is denoted by 
\eqn\STop{
\delta = \epsilon_+ Q_- - \epsilon_- Q_+ - \Bepsilon_+ \BQ_- + \Bepsilon_- \BQ_+.
}
The transformations for the vector superfield are written in the component fields \vects\ 
by \Wii
\eqna\STvec
$$\eqalignno{
&\delta \sigma = - i\sqrt{2}(\epsilon_-\Blambda_+ + \Bepsilon_+\lambda_-) ,&\STvec a\cr
&\delta \Bsigma = - i\sqrt{2}(\epsilon_+\Blambda_- + \Bepsilon_-\lambda_+) ,&\STvec b\cr
&\delta (v_0 - v_1) = 2i (\epsilon_-\Blambda_- + \Bepsilon_-\lambda_-) ,&\STvec c\cr
&\delta (v_0 + v_1) = 2i (\epsilon_+\Blambda_+ + \Bepsilon_+\lambda_+) ,&\STvec d\cr
&\delta D = \epsilon_+(\partial_0 - \partial_1)\Blambda_+
            + \epsilon_-(\partial_0 + \partial_1)\Blambda_-
            - \Bepsilon_+(\partial_0 - \partial_1)\lambda_+
            - \Bepsilon_-(\partial_0 + \partial_1)\lambda_- ,&\STvec e\cr
&\delta \lambda_+ = \sqrt{2}\epsilon_-(\partial_0 + \partial_1)\Bsigma 
                    - \epsilon_+(v_{01} - iD) ,&\STvec f\cr 
&\delta \lambda_- = \sqrt{2}\epsilon_+(\partial_0 - \partial_1)\sigma 
                    + \epsilon_-(v_{01} + iD) , &\STvec g\cr 
&\delta \Blambda_+ = \sqrt{2}\Bepsilon_-(\partial_0 + \partial_1)\sigma 
                    - \Bepsilon_+(v_{01} + iD) ,&\STvec h\cr 
&\delta \Blambda_- = \sqrt{2}\Bepsilon_+(\partial_0 - \partial_1)\Bsigma 
                    + \Bepsilon_-(v_{01} - iD) .&\STvec i
}$$
The transformations for the ordinary bosonic chiral superfield 
\bcs\ become
\eqna\STchib
$$\eqalignno{
\delta \phi^I &= \sqrt{2} (\epsilon_+ \psi^I_- - \epsilon_- \psi^I_+),&\STchib a\cr
\delta \psi^I_+ &= \sqrt{2}\epsilon_+F^I 
+ i\sqrt{2}\Bepsilon_-(D_0+D_1)\phi^I - 2Q_I\Bepsilon_+\Bsigma\phi^I ,&\STchib b\cr
\delta \psi^I_- &= \sqrt{2}\epsilon_-F^I
- i\sqrt{2}\Bepsilon_+(D_0-D_1)\phi^I + 2Q_I\Bepsilon_-\sigma\phi^I ,&\STchib c\cr
\delta F^I &= -i\sqrt{2}\Bepsilon_+(D_0-D_1)\psi^I_+ 
-i\sqrt{2}\Bepsilon_-(D_0+D_1)\psi^I_- \cr
&\phantom{=} +2Q_I(\Bepsilon_+\Bsigma\psi^I_-+\Bepsilon_-\sigma\psi^I_+)
-2iQ_I(\Bepsilon_+\Blambda_--\Bepsilon_-\Blambda_+)\phi^I ,&\STchib d
}$$
while the bosonic anti-chiral superfield 
\bacs\ are transformed as 
\eqna\STchiba
$$\eqalignno{
\delta \Bphi^I &= - \sqrt{2} (\Bepsilon_+ \Bpsi^I_- - \Bepsilon_- \Bpsi^I_+) , &\STchiba a\cr
\delta \Bpsi^I_+ &= \sqrt{2}\Bepsilon_+ \BF^I 
- i\sqrt{2}\epsilon_-(\BD_0+\BD_1)\Bphi^I - 2Q_I\epsilon_+\sigma\Bphi^I , &\STchiba b\cr
\delta \Bpsi^I_- &= \sqrt{2}\Bepsilon_- \BF^I
+ i\sqrt{2}\epsilon_+(\BD_0-\BD_1)\Bphi^I + 2Q_I\epsilon_-\Bsigma\Bphi^I , &\STchiba c\cr
\delta \BF^I &= -i\sqrt{2}\epsilon_+(\BD_0-\BD_1)\Bpsi^I_+ 
-i\sqrt{2}\epsilon_-(\BD_0+\BD_1)\Bpsi^I_- \cr
&\phantom{=} -2Q_I(\epsilon_+\sigma\Bpsi^I_-+\epsilon_-\Bsigma\Bpsi^I_+)
-2iQ_I(\epsilon_+\lambda_--\epsilon_-\lambda_+)\Bphi^I . &\STchiba d
}$$
We also calculate the supersymmetric transformations for 
the fermionic chiral superfield \fcs, 
\eqna\STchif
$$\eqalignno{
\delta \xi^A &= \sqrt{2} (\epsilon_+b^A_- - \epsilon_-b^A_+), &\STchif a\cr
\delta b^A_+ &= \sqrt{2}\epsilon_+\chi^A 
+ i\sqrt{2}\Bepsilon_-(D_0+D_1)\xi^A - 2q_A\Bepsilon_+\Bsigma\xi^A, &\STchif b\cr
\delta b^A_- &= \sqrt{2}\epsilon_-\chi^A
- i\sqrt{2}\Bepsilon_+(D_0-D_1)\xi^A + 2q_A\Bepsilon_-\sigma\xi^A , &\STchif c\cr
\delta \chi^A &= -i\sqrt{2}\Bepsilon_+(D_0-D_1)b^A_+ 
-i\sqrt{2}\Bepsilon_-(D_0+D_1)b^A_- \cr
&\phantom{=} +2q_A(\Bepsilon_+\Bsigma b^A_-+\Bepsilon_-\sigma b^A_+)
-2iq_A(\Bepsilon_+\Blambda_--\Bepsilon_-\Blambda_+)\xi^A. &\STchif d
}$$
These are the analogues of the bosonic chiral superfield \STchib{}.
The supersymmetric transformations for the fermionic anti-chiral superfield
\facs\ are given by
\eqna\STchifa
$$\eqalignno{
\delta \Bxi^A &= \sqrt{2} (\Bepsilon_+\Bb^A_- - \Bepsilon_-\Bb^A_+), &\STchifa a\cr
\delta \Bb^A_+ &= - \sqrt{2}\Bepsilon_+\Bchi^A 
+ i\sqrt{2}\epsilon_-(\BD_0+\BD_1)\Bxi^A + 2q_A\epsilon_+\sigma\Bxi^A, &\STchifa b\cr
\delta \Bb^A_- &= - \sqrt{2}\Bepsilon_-\Bchi^A 
- i\sqrt{2}\epsilon_+(\BD_0-\BD_1)\Bxi^A - 2q_A\epsilon_-\Bsigma\Bxi^A, &\STchifa c\cr
\delta \Bchi^A &= i\sqrt{2}\epsilon_+(\BD_0-\BD_1)\Bb^A_+ 
+i\sqrt{2}\epsilon_-(\BD_0+\BD_1)\Bb^A_- \cr
&\phantom{=} +2q_A(\epsilon_+\sigma \Bb^A_-+\epsilon_-\Bsigma \Bb^A_+)
-2iq_A(\epsilon_+\lambda_--\epsilon_-\lambda_+)\Bxi^A. &\STchifa d
}$$
In this calculation, we should note that, though the transformations \STchifa{}\ are the analogues 
of the ones \STchiba{}\ for the bosonic anti-chiral field, 
some signs in \STchifa{} are different from the ones in \STchiba{}.
It originates in whether the component fields  are 
commutative or anti-commutative.

Now we consider the world-sheet $\Sigma = \{(x^0,x^1)|x^1\geq 0\}$ 
with the boundary denoted by 
$\partial \Sigma = \{(x^0,x^1)|x^1=0\}$.
After the supersymmetric transformation for the Lagrangian \Lag, 
boundary terms are left on $\partial \Sigma$. 
In order for the theory to preserve the supersymmetry, 
it is necessary for the boundary terms to vanish. 
By using \STvec{}, \STchib{}, \STchiba{}, \STchif{}\ and \STchifa{}, 
let us calculate the variation of the Lagrangian \Lag\ under 
the supersymmetric transformation \STop. 
The variation of the kinetic term of the bosonic chiral superfield
becomes
\eqn\STLbc{\eqalign{
\delta L_{\rm kin}^b &= \sum_I \biggl[\epsilon_+ (\partial_0 - \partial_1)
   \biggl\{ {1 \over 2\sqrt{2}}\psi^I_-(\BD_0+\BD_1)\Bphi^I
   - {1 \over 2\sqrt{2}}\Bphi^I(D_0+D_1)\psi^I_- \cr
&\phantom{=}- {i \over \sqrt{2}}\Bpsi^I_+ F^I
   - iQ_I\psi^I_+\sigma\Bphi^I + Q_I\Blambda_+\Bphi^I\phi^I \biggr\} \cr
&\phantom{=}+ \epsilon_- (\partial_0 + \partial_1) 
   \biggl\{ -{1 \over 2\sqrt{2}}\psi^I_+(\BD_0-\BD_1)\Bphi^I 
   + {1 \over 2 \sqrt{2}}\Bphi^I (D_0-D_1)\psi^I_+ \cr
&\phantom{=}- {i \over \sqrt{2}}\Bpsi^I_- F^I
   + iQ_I\psi^I_-\Bsigma\Bphi^I + Q_I\Blambda_-\Bphi^I\phi^I \biggr\} \cr
&\phantom{=}+ \Bepsilon_+ (\partial_0 - \partial_1) 
   \biggl\{ -{1 \over 2\sqrt{2}}\Bpsi^I_-(D_0+D_1)\phi^I
   + {1 \over 2\sqrt{2}}\phi^I_-(\BD_0+\BD_1)\Bpsi^I_- \cr
&\phantom{=}- {i \over \sqrt{2}}\psi^I_+ \BF^I
   - iQ_I\Bpsi^I_+\Bsigma\phi^I - Q_I\lambda_+\Bphi^I\phi^I \biggr\} \cr
&\phantom{=}+ \Bepsilon_- (\partial_0 + \partial_1)
   \biggl\{ {1 \over 2\sqrt{2}}\Bpsi^I_+(D_0-D_1)\phi^I
   -{1 \over 2\sqrt{2}}\phi^I (\BD_0-\BD_1)\Bpsi^I_+ \cr 
&\phantom{=}- {i \over \sqrt{2}}\psi^I_- \BF^I
   + iQ_I\Bpsi^I_-\sigma\phi^I - Q_I\lambda_-\Bphi^I\phi^I \biggr\} 
\biggr] ,
}}
while that of the fermionic chiral superfield is
\eqn\STLfc{\eqalign{
\delta L_{\rm kin}^f &= \sum_A \biggl[\epsilon_+ (\partial_0 - \partial_1) 
   \biggl\{ -{1 \over 2\sqrt{2}}b^A_-(\BD_0+\BD_1)\Bxi^A 
   + {1 \over 2\sqrt{2}}\Bxi^A(D_0+D_1)b^A_- \cr
&\phantom{=}+ {i \over \sqrt{2}}\Bb^A_+\chi^A
   + iq_Ab^A_+\sigma\Bxi^A + q_A\Blambda_+\Bxi^A\xi^A \biggr\} \cr
&\phantom{=}+ \epsilon_- (\partial_0 + \partial_1) 
   \biggl\{ {1 \over 2\sqrt{2}}b^A_+(\BD_0-\BD_1)\Bxi^A 
   - {1 \over 2\sqrt{2}}\Bxi^A(D_0-D_1)b^A_+ \cr
&\phantom{=}+ {i \over \sqrt{2}}\Bb^A_-\chi^A
   - iq_Ab^A_-\Bsigma\Bxi^A + q_A\Blambda_-\Bxi^A\xi^A \biggr\} \cr
&\phantom{=}+ \Bepsilon_+ (\partial_0 - \partial_1) 
   \biggl\{ {1 \over 2\sqrt{2}}\Bb^A_-(D_0+D_1)\xi^A 
   -{1 \over 2\sqrt{2}}\xi^A(\BD_0+\BD_1)\Bb^A_- \cr
&\phantom{=}+ {i \over \sqrt{2}}b^A_+\Bchi^A
   + iq_A\Bb^A_+\Bsigma\xi^A - q_A\lambda_+\Bxi^A\xi^A \biggr\} \cr
&\phantom{=}+ \Bepsilon_- (\partial_0 + \partial_1) 
   \biggl\{ - {1 \over 2\sqrt{2}}\Bb^A_+(D_0-D_1)\xi^A 
   + {1 \over 2\sqrt{2}}\xi^A(\BD_0-\BD_1)\Bb^A_+ \cr
&\phantom{=}+ {i \over \sqrt{2}}b^A_-\Bchi^A
   - iq_A\Bb^A_-\sigma\xi^A - q_A\lambda_-\Bxi^A\xi^A \biggr\} \biggr] .
}}
The gauge kinetic term is transformed;
\eqn\STLg{\eqalign{
\delta L_{\rm gauge} &= {1 \over e^2} \biggl[
\epsilon_+ (\partial_0 - \partial_1) 
   \biggl\{ -{i \over 2\sqrt{2}}\Blambda_-(\partial_0+\partial_1)\sigma
   +{i \over 2\sqrt{2}}\sigma(\partial_0+\partial_1)\Blambda_- 
   + {1 \over 2}(D + iv_{01})\Blambda_+ \biggr\} \cr
&\phantom{=}+ \epsilon_- (\partial_0 + \partial_1) 
   \biggl\{ -{i \over 2\sqrt{2}}\Blambda_+(\partial_0-\partial_1)\Bsigma
   +{i \over 2\sqrt{2}}\Bsigma(\partial_0-\partial_1)\Blambda_+ 
   + {1 \over 2}(D - iv_{01})\Blambda_- \biggr\} \cr
&\phantom{=}+ \Bepsilon_+ (\partial_0 - \partial_1) 
   \biggl\{ -{i \over 2\sqrt{2}}\lambda_-(\partial_0+\partial_1)\Bsigma
   +{i \over 2\sqrt{2}}\Bsigma(\partial_0+\partial_1)\lambda_- 
   - {1 \over 2}(D - iv_{01})\lambda_+ \biggr\} \cr
&\phantom{=}+ \Bepsilon_- (\partial_0 + \partial_1) 
   \biggl\{ -{i \over 2\sqrt{2}}\lambda_+(\partial_0-\partial_1)\sigma
   +{i \over 2\sqrt{2}}\sigma(\partial_0-\partial_1)\lambda_+ 
   - {1 \over 2}(D + iv_{01})\lambda_- \biggr\} \biggr] .
}}
The susy variation for the FI and theta angle terms is written down as 
\eqn\STLdt{\eqalign{
\delta L_{D,\theta} &= 
\epsilon_+\biggl(-r + i {\theta \over 2\pi}\biggr)(\partial_0-\partial_1)\Blambda_+ 
+\epsilon_-\biggl(-r - i {\theta \over 2\pi}\biggr)(\partial_0+\partial_1)\Blambda_- \cr
&\phantom{=}+\Bepsilon_+\biggl(r + i {\theta \over 2\pi}\biggr)(\partial_0-\partial_1)\lambda_+ 
+\Bepsilon_-\biggl(r - i {\theta \over 2\pi}\biggr)(\partial_0+\partial_1)\lambda_- .
}}
Using the left derivative for $\partial /\partial \xi$ and 
$\partial / \partial \Bxi$, we can represent
the transformation for the superpotential as
\eqn\STLsupot{\eqalign{
\delta L_W &= i \sqrt{2} \biggl[
\epsilon_+ (\partial_0 - \partial_1) 
    \biggl(\sum_I\Bpsi_+^I{\partial \BW(\Bphi,\Bxi) \over \partial \Bphi^I}
	-\sum_A \Bb_+^A{\partial \BW(\Bphi,\Bxi) \over \partial \Bxi^A}\biggr) \cr
&\phantom{=}+\epsilon_- (\partial_0 + \partial_1) 
    \biggl(\sum_I\Bpsi_-^I{\partial \BW(\Bphi,\Bxi) \over \partial \Bphi^I}
	-\sum_A \Bb_-^A{\partial \BW(\Bphi,\Bxi) \over \partial \Bxi^A}\biggr) \cr
&\phantom{=}+\Bepsilon_+ (\partial_0 - \partial_1) 
    \biggl(\sum_I\psi_+^I{\partial W(\phi,\xi) \over \partial \phi^I}
	+\sum_A b_+^A{\partial W(\phi,\xi) \over \partial \xi^A}\biggr) \cr
&\phantom{=}+\Bepsilon_- (\partial_0 + \partial_1) 
    \biggl(\sum_I\psi_-^I{\partial W(\phi,\xi) \over \partial \phi^I}
	+\sum_A b_-^A{\partial W(\phi,\xi) \over \partial \xi^A}\biggr) 
\biggr] .
}}
Following the supersymmetric transformation \STop, we decompose 
the boundary terms induced by the variation of the action 
$S = \int_\Sigma d^2x L$ as 
\eqn\STLag{
\delta S = -\int_{\partial\Sigma} dx^0 
(\epsilon_+ \CL^+ - \epsilon_- \CL^- 
- \Bepsilon_+ \BCL^+ + \Bepsilon_- \BCL^-) .
}
The components $\CL^\pm$ and $\BCL^\pm$ 
are summarised in appendix A, 
where the auxiliary fields $F$, $\chi$ and $D$ are explicitly included.
Integrating out these auxiliary fields in terms of the equations of motion
\eomaux{}, we obtain 
\eqn\bndp{\eqalign{
\CL^+ = &-\sum_I\biggl({1 \over 2\sqrt{2}}\psi^I_-(\BD_0+\BD_1)\Bphi^I
   - {1 \over 2\sqrt{2}}\Bphi^I(D_0+D_1)\psi^I_- 
   - iQ_I\psi^I_+\sigma\Bphi^I \biggr) \cr
   &+\sum_A\biggl({1 \over 2\sqrt{2}}b^A_-(\BD_0+\BD_1)\Bxi^A 
   - {1 \over 2\sqrt{2}}\Bxi^A(D_0+D_1)b^A_- 
   - iq_Ab^A_+\sigma\Bxi^A \biggr) \cr
   &+ {i \over 2\sqrt{2}e^2}\bigl( \Blambda_-(\partial_0+\partial_1)\sigma
   - \sigma(\partial_0+\partial_1)\Blambda_- \bigr) 
   - T \Blambda_+ \cr
   &-{i \over \sqrt{2}}\biggl(\sum_I\Bpsi_+^I{\partial \BW(\Bphi,\Bxi) \over \partial \Bphi^I}
   -\sum_A\Bb_+^A{\partial \BW(\Bphi,\Bxi) \over \partial \Bxi^A}\biggr) ,
}}
\eqn\bndm{\eqalign{
\CL^- = &\sum_I\biggl({1 \over 2\sqrt{2}}\psi^I_+(\BD_0-\BD_1)\Bphi^I 
   - {1 \over 2 \sqrt{2}}\Bphi^I (D_0-D_1)\psi^I_+ 
   - iQ_I\psi^I_-\Bsigma\Bphi^I \biggr) \cr
   &- \sum_A\biggl({1 \over 2\sqrt{2}}b^A_+(\BD_0-\BD_1)\Bxi^A 
   - {1 \over 2\sqrt{2}}\Bxi^A(D_0-D_1)b^A_+ 
   - iq_Ab^A_-\Bsigma\Bxi^A \biggr) \cr
   &+ {i \over 2\sqrt{2}e^2}\bigl( \Blambda_+(\partial_0-\partial_1)\Bsigma
   - \Bsigma(\partial_0-\partial_1)\Blambda_+ \bigr) 
   - \BT \Blambda_- \cr
   &- {i \over \sqrt{2}}\biggl(\sum_I\Bpsi_-^I{\partial \BW(\Bphi,\Bxi) \over \partial \Bphi^I}
   - \sum_A\Bb_-^A{\partial \BW(\Bphi,\Bxi) \over \partial \Bxi^A}\biggr) ,
}}
\eqn\bndpb{\eqalign{
\BCL^+ = &- \sum_I\biggl({1 \over 2\sqrt{2}}\Bpsi^I_-(D_0+D_1)\phi^I
   - {1 \over 2\sqrt{2}}\phi^I(\BD_0+\BD_1)\Bpsi^I_- 
   + iQ_I\Bpsi^I_+\Bsigma\phi^I \biggr) \cr
   &+\sum_A\biggl({1 \over 2\sqrt{2}}\Bb^A_-(D_0+D_1)\xi^A 
   -{1 \over 2\sqrt{2}}\xi^A(\BD_0+\BD_1)\Bb^A_- 
   + iq_A\Bb^A_+\Bsigma\xi^A \biggr) \cr
   &- {i \over 2\sqrt{2}e^2}\bigl(\lambda_-(\partial_0+\partial_1)\Bsigma
   - \Bsigma(\partial_0+\partial_1)\lambda_- \bigl) 
   - \BT\lambda_+ \cr
   &+ {i \over \sqrt{2}}\biggl(\sum_I\psi_+^I{\partial W(\phi,\xi) \over \partial \phi^I}
   + \sum_A b_+^A{\partial W(\phi,\xi) \over \partial \xi^A}\biggr) ,
}}
\eqn\bndmb{\eqalign{
\BCL^- = &\sum_I\biggl({1 \over 2\sqrt{2}}\Bpsi^I_+(D_0-D_1)\phi^I
   -{1 \over 2\sqrt{2}}\phi^I (\BD_0-\BD_1)\Bpsi^I_+ 
   + iQ_I\Bpsi^I_-\sigma\phi^I \biggr) \cr
   &-\sum_A \biggl({1 \over 2\sqrt{2}}\Bb^A_+(D_0-D_1)\xi^A 
   - {1 \over 2\sqrt{2}}\xi^A(\BD_0-\BD_1)\Bb^A_+ 
   + iq_A\Bb^A_-\sigma\xi^A \biggr) \cr
   &- {i \over 2\sqrt{2}e^2}\bigl( \lambda_+(\partial_0-\partial_1)\sigma
   - \sigma(\partial_0-\partial_1)\lambda_+ \bigr) 
   -T\lambda_- \cr
   &+{i \over \sqrt{2}} \biggl(\sum_I\psi_-^I{\partial W(\phi,\xi) \over \partial \phi^I}
   + \sum_A b_-^A{\partial W(\phi,\xi) \over \partial \xi^A}\biggr) , 
}}
where $T$ is defined as
$$
T \equiv {1 \over 2}\biggl(\sum_I Q_I \Bphi^I\phi^I +\sum_A q_A \Bxi^A\xi^A - r\biggr)
+ i \biggl({v_{01} \over 2e^2} + {\theta \over 2\pi}\biggr) .
$$
We can consider two types of the supersymmetric boundary conditions. 
One is called the A-type boundary condition, which is defined by 
$$
\epsilon_+ = \eta \Bepsilon_-,\quad \epsilon_- = \eta \Bepsilon_+,  
$$
where $\eta \equiv \pm 1$.
This boundary condition is then satisfied by 
\eqn\Abndcond{
\eta \CL^+ + \BCL^- = 0,\quad 
\eta \CL^- + \BCL^+ = 0\quad \hbox{on $\partial \Sigma$}. 
}
The other is the B-type boundary condition defined by 
$$
\epsilon_+ = \eta\epsilon_-,\quad \Bepsilon_+ = \eta\Bepsilon_- ,
$$
and the B-type boundary condition is realised when  
\eqn\Bbndcond{
\eta \CL^+ - \CL^- = 0,\quad \eta \BCL^+ - \BCL^- = 0 \quad\hbox{on $\partial \Sigma$}.
}
When we look at the fields $\lambda_\pm$, $\Blambda_\pm$, $\sigma$ and $\Bsigma$ in \Abndcond\ and \Bbndcond, 
we can read the boundary conditions for the 
vector superfield; Dirichlet type on $\partial\Sigma$ for 
the A-type boundary. 
For the B-type case, when we switch off the theta parameter $\theta=0$, 
$\Sigma=\BSigma$ on $\partial \Sigma$ 
is obtained.  
Especially these mean $\lambda_+ + \eta\Blambda_- =0$ and 
$\lambda_- + \eta \Blambda_+=0$ for the A-type case. 
On the other hand, the B-type case leads us to 
$\lambda_+-\eta\lambda_- =0$ and $\Blambda_+ - \eta \Blambda =0$ 
with $\theta =0$. 

The loci of $\phi$ and $\xi$ satisfying \Abndcond\ and \Bbndcond\ are 
regarded as the A-brane and the B-brane respectively. 
In the ordinary sigma model with only the bosonic chiral superfields, 
such D-branes were analysed in the $e^2 \to \infty$ limit in \refs{\GJS,\GJSi}.

\subsec{$e^2 \to \infty$ limit}

We shall consider the boundary conditions in the $e^2 \to \infty$ limit 
of the gauged linear sigma model on the supermanifold.
In this limit, dynamics of fields in the gauge multiplet is fixed by 
equations of motion and this theory is reduced to the 
nonlinear sigma model, so that 
the geometry of the target space can be analysed. 
The equations of motion for 
$\lambda_\pm$ and $\Blambda_\pm$, 
are calculated as
\eqn\eomlamL{\eqalign{
\sum_IQ_I\Bphi^I\psi_-^I - \sum q_A\Bxi^Ab_-^A = 0,\quad 
\sum_IQ_I\Bphi^I\psi_+^I - \sum q_A\Bxi^Ab_+^A = 0, \cr
\sum_IQ_I\phi^I\Bpsi_-^I - \sum q_A\xi^A\Bb_-^A = 0,\quad
\sum_IQ_I\phi^I\Bpsi_+^I - \sum q_A\xi^A\Bb_+^A = 0. 
}}
If we introduce 
$K(\phi,\xi) \equiv \sum_IQ_I^2\Bphi^I\phi^I + \sum_Aq_A^2\Bxi^A\xi^A$, 
the equations of motion for $\sigma$ and $ \Bsigma$ are described as 
\eqn\eomsigL{
\Bsigma = - {\sum_IQ_I\Bpsi_-^I\psi_+^I + \sum_Aq_A\Bb_-^Ab_+^A 
\over \sqrt{2}K(\phi,\xi)} ,\quad
\sigma = - {\sum_IQ_I\Bpsi_+^I\psi_-^I + \sum_Aq_A\Bb_+^Ab_-^A 
\over \sqrt{2}K(\phi,\xi)} .
}
Substituting these equations into the interaction terms
$-2\Bsigma\sigma\biggl(\sum_IQ_I^2\Bphi^I\phi^I + \sum_Aq_A^2\Bxi^A\xi^A\biggr)$
in the Lagrangian $L^b_{\rm kin} + L^f_{\rm kin}$,
we obtain expression of the interaction terms 
$$\eqalign{
-{1 \over K(\phi,\xi)}
\biggl[\sum_{I,J}Q_IQ_J\Bpsi_-^I\psi_+^I\Bpsi_+^J\psi_-^J 
+ \sum_{A,B}q_Aq_B\Bb_-^Ab_+^A\Bb_+^Bb_-^B \cr
+ \sum_{I,A}Q_Iq_A\Bigl(\Bpsi_-^I\psi_+^I\Bb_+^Ab_-^A
+ \Bpsi_+^I\psi_-^I\Bb_-^Ab_+^A\Bigr)
\biggr] .
}$$
These terms are the analogues of the four fermi interaction term 
$R_{{\bar i}j{\bar k}l}\psi^{\bar i}\psi^j\psi^{\bar k}\psi^l$ 
in the nonlinear sigma model on the ordinary bosonic manifold.
We can also evaluate the equations of motion for the gauge fields $v_0$ and $v_1$ 
in this limit; 
\eqn\eomvL{\eqalign{
2v_0K(\phi,\xi)
= &\sum_IQ_I\Bigl[i\Bigl(\Bphi^I\partial_0\phi^I - \phi^I\partial_0\Bphi^I\Bigr)
      + \Bpsi_+^I\psi_+^I + \Bpsi_-^I\psi_-^I\Bigr] \cr
&+ \sum_Aq_A\Bigl[i\Bigl(\Bxi^A\partial_0\xi^A + \xi^A\partial_0\Bxi^A\Bigr)
      + \Bb_+^Ab_+^A + \Bb_-^Ab_-^A\Bigr] , \cr
2v_1K(\phi,\xi)
=& \sum_IQ_I\Bigl[i\Bigl(\Bphi^I\partial_1\phi^I - \phi^I\partial_1\Bphi^I\Bigr)
      + \Bpsi_+^I\psi_+^I - \Bpsi_-^I\psi_-^I\Bigr] \cr
&+ \sum_Aq_A\Bigl[i\Bigl(\Bxi^A\partial_1\xi^A + \xi^A\partial_1\Bxi^A\Bigr)
      + \Bb_+^Ab_+^A - \Bb_-^Ab_-^A\Bigr] .
}}

Now let us consider the A-type and the B-type boundary conditions, 
which are denoted by \Abndcond\ and \Bbndcond\ respectively, 
in the $e^2 \to \infty$ limit. 

\medskip
\noindent $\underline{\hbox{\it A-type boundary condition}}$
\medskip
Substituting \bndp, \bndm, \bndpb\ and \bndmb\ into \Abndcond, 
the A-type boundary condition is read as
\eqna\eiAbnd
$$\eqalignno{
&\sum_I \Bigl(- \eta \psi_-^I (\BD_0+\BD_1) \Bphi^I 
		+ \Bpsi_+^I (D_0-D_1) \phi^I \Bigr) 
+ \sum_A \Bigl(\eta b_-^A (\BD_0+\BD_1) \Bxi^A 
		- \Bb_+^A (D_0-D_1) \xi^A \Bigr) \cr
&= \sqrt{2}T (\lambda_- + \eta \Blambda_+)
- i \biggl[
\sum_I\biggl(\psi_-^I{\partial W \over \partial \phi^I} 
            - \eta \Bpsi_+^I{\partial \BW \over \partial \Bphi^I}\biggr)
+ \sum_A \biggl(b_-^A{\partial W \over \partial \xi^A} 
            + \eta \Bb_+^A{\partial \BW \over \partial \Bxi^A}\biggr)\biggr] , &\eiAbnd{a} \cr
&\sum_I \Bigl(- \eta\psi_+^I (\BD_0-\BD_1) \Bphi^I 
			+ \Bpsi_-^I (D_0+D_1) \phi^I \Bigr) 
+ \sum_A \Bigl(\eta b_+^A (\BD_0-\BD_1) \Bxi^A 
			- \Bb_-^A (D_0+D_1) \xi^A \Bigr) \cr
&= -\sqrt{2} \BT (\lambda_+ + \eta \Blambda_-) 
+i\biggl[\sum_I\biggl(\psi_+^I {\partial W \over \partial \phi^I} 
                 - \eta\Bpsi_-^I {\partial \BW \over \partial \Bphi^I}\biggr) 
+ \sum_A\biggl(b_+^A {\partial W \over \partial \xi^A} 
                 + \eta \Bb_-^A {\partial \BW \over \partial \Bxi^A}\biggr)\biggr] . &\eiAbnd{b} \cr
}$$
We shall show a simple solution satisfying 
the A-type boundary condition \eiAbnd{}. 
We assume that on the boundary $\partial \Sigma$
\eqn\Abdrl{
\lambda_+ + \eta \Blambda_- = 0,\quad \lambda_- + \eta \Blambda_+ = 0 , 
}
and put the boundary conditions for $\psi^I$ and $b^A$, 
\eqn\Abdrp{
\eta \psi_-^I - \Bpsi_+^I = 0,\quad \eta b_-^A - \Bb_+^A = 0 .
}
They lead to the constraints for the superpotential on $\partial \Sigma$, 
\eqn\AbdrW{
{\partial W \over \partial \phi^I} - {\partial \BW \over \partial \Bphi^I} = 0, \quad 
{\partial W \over \partial \xi^A} + {\partial \BW \over \partial \Bxi^A} = 0 .
}
By the use of \eomlamL, \Abdrl, \Abdrp\ and \AbdrW, the gauge fields 
are eliminated from the boundary conditions \eiAbnd{} and these conditions are reduced to 
\eqna\redAbnd
$$\eqalignno{
\sum_I \Bigl[\eta\psi_-^I(\partial_0+\partial_1)\Bphi^I 
- \Bpsi_+^I(\partial_0-\partial_1)\phi^I\Bigr]
- \sum_A \Bigl[\eta b_-^A(\partial_0+\partial_1)\Bxi^A 
- \Bb_+^A(\partial_0-\partial_1)\xi^A\Bigr] = 0 , &\redAbnd a\cr
\sum_I \Bigl[\eta\psi_+^I(\partial_0-\partial_1)\Bphi^I 
- \Bpsi_-^I(\partial_0+\partial_1)\phi^I\Bigr]
- \sum_A \Bigl[\eta b_+^A(\partial_0-\partial_1)\Bxi^A 
- \Bb_-^A(\partial_0+\partial_1)\xi^A\Bigr] = 0 . &\redAbnd b
}$$
Under the constraint \Abdrp\
, the equations \redAbnd{} can be satisfied
by $(\partial_0+\partial_1)\Bphi = (\partial_0-\partial_1)\phi$ and 
$(\partial_0+\partial_1)\Bxi = (\partial_0-\partial_1)\xi$. 
Namely, we can take 
\eqn\bAbrane{
\partial_0 (\Im \phi^I)=0,
\quad \partial_1 (\Re \phi^I) = 0,
}
for the bosonic coordinates $\phi^I$ in the target space, and 
\eqn\fAbrane{
\partial_0 (\Im \xi^A) =0,
\quad \partial_1 (\Re \xi^A) = 0,
}
for the fermionic coordinates $\xi^A$. From \bAbrane, 
the Dirichlet boundary conditions are imposed on 
real $m$-dimensional boundaries in the complex $m$-dimensional 
bosonic part of the target supermanifold. 
This bosonic part of D-brane with half of the dimensions 
of the target manifold should correspond 
to the A-brane on the Lagrangian submanifold 
in the ordinary sigma model. 
In the same way, 
since the Dirichlet boundary conditions are imposed on the boundaries 
with half of the dimensions of the Grassmann odd part 
of the target supermanifold, 
\fAbrane\ implies the existence of the fermionic A-brane. 

We discussed the simple condition 
denoted by \Abdrp, \AbdrW, \bAbrane\ and \fAbrane\ .
In the setting of our model, there are indices of flavours ``$I$" and ``$A$". 
If several $U(1)$ charges $Q_I$'s and  $q_A$'s have a same value, 
we can mix the corresponding fields $\psi^I$ and $\xi^A$.  
If we put $Q_I = q_A$ for all $I$ and $A$, 
then we can present a more general boundary condition 
on the boundary $\partial \Sigma$;
$$\eqalign{
\eta \psi_-^I = M^I{}_J \Bpsi_+^J + F^I{}_B\Bb_+^B,&\quad 
\eta b_-^A = G^A{}_J \Bpsi_+^J + N^A{}_B \Bb_+^B, \cr
{\partial W \over \partial \phi^J}M^J{_I} 
+ {\partial W \over \partial \xi^A}G^A{}_I 
- {\partial \BW \over \partial \Bphi^I} = 0, &\quad
{\partial W \over \partial \phi^I}F^I{}_A 
+ {\partial W \over \partial \xi^B} N^B{}_A
+ {\partial \BW \over \partial \Bxi^A} = 0 , \cr
\partial_0 (\phi_I - \Bphi_JM^J{}_I + \Bxi_A G^A{}_I) = 0,&\quad 
\partial_1 (\phi_I + \Bphi_JM^J{}_I - \Bxi_A G^A{}_I) = 0, \cr
\partial_0 (\xi_A + \Bphi_JF^J{}_A - \Bxi_B N^B{}_A) = 0,&\quad 
\partial_1 (\xi_A - \Bphi_JF^J{}_A + \Bxi_B N^B{}_A) = 0.
}$$
$M$ and $N$ are Grassmann even, while $F$ and $G$ are Grassmann odd.
The previous discussion corresponds to 
the case with $M^I{}_J=\delta^I{}_J$, $N^A{}_B=\delta^A{}_B$ and  
$F^I{}_B=G^A{}_J=0$. 

\medskip
\noindent $\underline{\hbox{\it B-type boundary condition}}$
\medskip
First we consider B-type boundary conditions with the theta parameter $\theta=0$. 
By the use of \bndp, \bndm, \bndpb\ and \bndmb, the B-type boundary conditions 
\Bbndcond\ are rewritten in the $e^2 \to \infty$ limit, 
\eqna\eiBbnd
$$\eqalignno{
&\sum_I \Bigl(\eta \psi_-^I (\BD_0+\BD_1) \Bphi^I 
			+ \psi_+^I (\BD_0-\BD_1) \Bphi^I \Bigr) 
+ \sum_A \Bigl(-\eta b_-^A (\BD_0+\BD_1) \Bxi^A 
			- b_+^A (\BD_0-\BD_1) \Bxi^A \Bigr) \cr
&= - i \biggl[ \sum_I\bigl(\eta\Bpsi_+^I - \Bpsi_-^I \bigr) {\partial \BW \over \partial \Bphi^I} 
-\sum_A\bigl(\eta\Bb_+^A - \Bb_-^A\bigr){\partial \BW \over \partial \Bxi^A} 
\biggr] , &\eiBbnd{a} \cr
&\sum_I \Bigl(-\eta\Bpsi_-^I (D_0+D_1) \phi^I 
			- \Bpsi_+^I (D_0-D_1) \phi^I \Bigr) 
+ \sum_A \Bigl(\eta\Bb_-^A (D_0+D_1) \xi^A 
			+ \Bb_+^A (D_0-D_1) \xi^A \Bigr) \cr
&= - i \biggl[ \sum_I\bigl(\eta\psi_+^I - \psi_-^I \bigr) {\partial W \over \partial \phi^I} 
+\sum_A\bigl(\eta b_+^A - b_-^A\bigr){\partial W \over \partial \xi^A} 
\biggr] . &\eiBbnd{b} \cr
}$$
In order to show a simple solution, we assume that 
$$
\eta\lambda_+ - \lambda_- =0 ,\quad \eta\Blambda_+ - \Blambda_- = 0. 
$$
Then the conditions \eiBbnd{} are reduced by \eomlamL\ to 
\eqna\redBbnd
$$\eqalignno{
&\phantom{=} \sum_I \Bigl[(\psi_+^I + \eta \psi_-^I)\partial_0\Bphi^I 
- (\psi_+^I - \eta \psi_-^I)\partial_1\Bphi^I\Bigr]
- \sum_A \Bigl[(b_+^A +\eta b_-^A)\partial_0\Bxi^A 
- (b_+^A - \eta b_-^A)\partial_1\Bxi^A \Bigr] \cr
&= -i \biggl[\sum_I(\eta \Bpsi_+^I - \Bpsi_-^I){\partial \BW \over \partial \Bphi^I}
- \sum_A(\eta \Bb_+^A - \Bb_-^A){\partial \BW \over \partial \Bxi^A}\biggr] , &\redBbnd a \cr
&\phantom{=} \sum_I \Bigl[(\Bpsi_+^I + \eta \Bpsi_-^I)\partial_0\phi^I 
-(\Bpsi_+^I - \eta \Bpsi_-^I)\partial_1\phi^I\Bigr]
- \sum_A \Bigl[(\Bb_+^A + \eta \Bb_-^A)\partial_0\xi^A 
-(\Bb_+^A - \eta \Bb_-^A)\partial_1\xi^A \Bigr] \cr
&= i \biggl[\sum_I(\eta \psi_+^I - \psi_-^I){\partial W \over \partial \phi^I}
+ \sum_A(\eta b_+^A - b_-^A){\partial W \over \partial \xi^A}\biggr]. &\redBbnd b
}$$
We can choose the boundary condition, 
$\psi_+^I + \eta \psi_-^I = 0$ or $\psi_+^I - \eta \psi_-^I = 0$, 
in the bosonic part. When $\psi_+^I + \eta \psi_-^I = 0$, 
the conditions  for $\phi$ and $W$ are led to
\eqn\bosN{\partial_1 \phi^I = \partial_1 \Bphi^I = 0,\quad
{\partial W \over \partial \phi^I} = {\partial \BW \over \partial \Bphi^I} = 0.
}
These are interpreted as Neumann boundary condition.
On the other hand, if $\psi_+^I - \eta \psi_-^I = 0$, we then obtain 
on the boundary $\partial \Sigma$, 
\eqn\bosD{
\partial_0 \phi^I = \partial_0 \Bphi^I = 0, 
}
and this corresponds to Dirichlet boundary condition. 
In the same way, we can consider the fermionic part of \redBbnd{}. 
When we impose $b_+^A + \eta b_-^A = 0$, the boundary condition becomes 
\eqn\ferN{
\partial_1\xi^A = \partial_1\Bxi^A = 0,\quad 
{\partial W \over \partial \xi^A} = {\partial \BW \over \partial \Bxi^A} = 0, 
}
and these correspond to Neumann boundary condition.
When $b_+^A - \eta b_-^A = 0$, we obtain 
the Dirichlet boundary condition; 
\eqn\ferD{
\partial_0 \xi^I = \partial_0 \Bxi^I = 0. 
}

As mentioned above, we can impose Dirichlet or Neumann boundary condition on 
each bosonic and fermionic coordinate. 
Since both of the real and the imaginary parts 
of $\phi$ and $\xi$ must satisfy the same condition
from \bosD\ and \ferD, 
there exists a B-brane which should be a D-brane wrapping 
on the complex submanifold in the target supermanifold. 
In other words, by imposing the Dirichlet condition on appropriate 
bosonic and fermionic coordinates, we can obtain  
fermionic B-branes and super-B-branes as well as 
the bosonic B-branes on the subsupermanifold. 
This subsupermanifold might be regarded as a supersymmetric cycle 
in the supermanifold. 

We can also write down a general condition for the B-type boundary. 
We introduce Grassmann even matrices, ${\tilde M}$ and ${\tilde N}$, 
and Grassmann odd matrices, ${\tilde F}$ and ${\tilde G}$. 
The B-type boundary condition is reduced to 
\eqn\genBbdr{\eqalign{
\eta \psi_-^I = {\tilde M}^I{}_J \psi_+^J + {\tilde F}^I{}_B b_+^B, &\quad 
\eta b_-^A = {\tilde G}^A{}_J\psi_+^J + {\tilde N}^A{}_B b_+^B , \cr
\partial_0 [\phi_J(\delta^J{}_I + {\tilde M}^{*J}{}_I) 
+ \xi_A {\tilde G}^{*A}{}_I] = 0, &\quad
\partial_0 [\xi_B(\delta^B{}_A + {\tilde N}^{*B}{}_A) 
- \phi_I {\tilde F}^{*I}{}_A] = 0, \cr
\partial_1 [\phi_J(\delta^J{}_I - {\tilde M}^{*J}{}_I) 
- \xi_A {\tilde G}^{*A}{}_I] = 0, &\quad
\partial_1 [\xi_B(\delta^B{}_A - {\tilde N}^{*B}{}_A) 
+ \phi_I {\tilde F}^{*I}{}_A] = 0, \cr
{\partial W \over \partial \phi^J}(\delta^J{}_I - {\tilde M}^J{}_I) 
+ {\partial W \over \partial \xi^A} {\tilde G}^A{}_I = 0, &\quad
{\partial W \over \partial \xi^B}(\delta^B{}_A - {\tilde N}^B{}_A) 
- {\partial W \over \partial \phi^I} {\tilde F}^I{}_A = 0 .
}}
When ${\tilde M} = {\tilde N} = - {\bf 1}$, 
the Neumann boundary conditions \bosN\ and \ferN\ are reproduced. 
On the other hand, the Dirichlet boundary conditions 
\bosD\ and \ferD\ are reproduced when ${\tilde M} = {\tilde N} = {\bf 1}$. 
The general B-type boundary condition \genBbdr\ implies the existence of 
the D-branes embedded transversally in the supermanifold. 

Finally we make a comment on the role of the theta parameter 
in the B-type case.
By the use of \eomvL, the field strength 
$v_{01}=\partial_0v_1-\partial_1v_0$ is written as 
$$\eqalign{ 
v_{01} &=
{i\over K}\sum_IQ_I(D^B_0\Bphi^ID^B_1\phi^I-D^B_1\Bphi^ID^B_0\phi^I)
+{i\over K}\sum_Aq_A(D^B_0\Bxi^AD^B_1\xi^A-D^B_1\Bxi^AD^B_0\xi^A)\cr
&\phantom{=} +(\partial_0-\partial_1)\biggl[
             {1\over K}\biggl(\sum_IQ_I\Bpsi^I_+\psi^I_+ 
             +\sum_Aq_Ab^A_+b^A_+\biggr)\biggr] \cr
&\phantom{=} -(\partial_0+\partial_1)\biggl[
             {1\over K}\biggl(\sum_IQ_I\Bpsi^I_-\psi^I_- 
             +\sum_Aq_Ab^A_-b^A_-\biggr)\biggr].\cr
}$$
Here we introduced a new vector potential $v^B_i$ 
and its covariant derivatives given by 
$$\eqalign{
2v^B_0K(\phi,\xi)
&= \sum_IiQ_I\Bigl(\Bphi^I\partial_0\phi^I - \phi^I\partial_0\Bphi^I\Bigr)
+ \sum_Aiq_A\Bigl(\Bxi^A\partial_0\xi^A + \xi^A\partial_0\Bxi^A\Bigr), \cr
2v^B_1K(\phi,\xi)
&= \sum_IiQ_I\Bigl(\Bphi^I\partial_1\phi^I - \phi^I\partial_1\Bphi^I\Bigr)
+ \sum_Aiq_A\Bigl(\Bxi^A\partial_1\xi^A + \xi^A\partial_1\Bxi^A\Bigr),\cr
D^B_i\phi^I &= (\partial_i +iQ_Iv^B_i)\phi^I,\quad 
D^B_i\xi^A = (\partial_i+iq_Av^B_i)\xi^A .\cr
}$$
This field strength $v_{01}$ induces the theta-term $L_{\theta}$ and 
boundary theta-term $L_{{\rm bdy},\theta}$
$$\eqalign{
S_{\theta}&=\int_{\Sigma}d^2x L_{\theta}+\int_{\partial\Sigma}dx^0 L_{{\rm bdy},\theta},\cr
L_{\theta}&= {\theta\over {2\pi}}\biggl[
{i\over K}\sum_I Q_I \bigl(D^B_0 \Bphi^ID^B_1\phi^I-D^B_1\Bphi^ID^B_0\phi^I\bigr) \cr
&\phantom{= {\theta\over {2\pi}}\biggl[} +{i\over K}
\sum_Aq_A \bigl(D^B_0\Bxi^A D^B_1\xi^A-D^B_1\Bxi^A D^B_0 \xi^A\bigr)
\biggr], \cr
L_{{\rm bdy},\theta}&=
{\theta\over {2\pi}}\biggl[
{1\over K}\biggl(\sum_IQ_I\Bpsi^I_+ \psi^I_+ +\sum_Aq_Ab^A_+b^A_+\biggr)
+{1\over K}\biggl(\sum_IQ_I\Bpsi^I_-\psi^I_- +\sum_Aq_Ab^A_-b^A_-\biggr)
\biggr].\cr
}$$
Then $L_{\theta}$ is interpreted as an effect of an antisymmetric tensor field $B$. 
This field is combined with the K{\" a}hler form and 
complexifies the K{\" a}hler parameter of the 
nonlinear sigma model.
The boundary term $L_{{\rm bdy},\theta}$ is interpreted as a kind of 
(supersymmetric version of) ${\theta\over {2\pi}}\int_{\partial\Sigma}dx^0 v_0$.
When we take B-type boundary condition 
on $\partial\Sigma$, 
$L_{{\rm bdy},\theta}$ is represented by $\sigma$ and $\Bsigma$; 
$
S_{{\rm bdy},\theta}={\theta\over {2\pi}}
\int_{\partial\Sigma}dx^0{\eta\over \sqrt{2}}(\sigma + \Bsigma) .
$
That is consistent with the fact that 
the vector potential $V$ is reduced to 
$V=2\theta\Btheta(v_0+{\eta\over\sqrt{2}}(\sigma +\Bsigma))$ on 
the boundary.

Next we shall study the effect of the theta parameter in the context of 
gauged linear sigma model. 
For simplicity, we put the potential $W=0$. 
The bulk action changes under the supersymmetric transformation. 
But the resulting boundary terms in the B-type can be cancelled by 
adding boundary action $S_{\rm bdy}=\int_{\partial\Sigma} dx^0 L_{\rm bdy}$
to the bulk action $S_{\rm bulk}$. We introduce  
total action $S_{\rm tot}$ by combining $S_{\rm bulk}$ and $S_{\rm bdy}$
$$\eqalign{
S_{\rm tot}&=\int_{\Sigma}d^2x L_{\rm bulk}
		+\int_{\partial\Sigma}dx^0 L_{\rm bdy}, \cr
L_{\rm bdy}&=
{1\over 4}\partial_1\biggl(\sum_I\Bphi^I \phi^I+\sum_A\Bxi^A \xi^A 
-{1\over e^2}\Bsigma\sigma\biggr)\cr
&\phantom{=} +{i\eta\over 2}\sum_I\bigl(\Bpsi_-^I \psi_+^I 
- \Bpsi_+^I \psi_-^I \bigr)
+{i\eta\over 2}\sum_A\bigl(\Bb_-^A b_+^A - \Bb_+^A b_-^A\bigr)
-{i\eta\over \sqrt{2}}(T\sigma -\BT\Bsigma) . \cr
}$$
Then $S_{\rm tot}$ is invariant under the supersymmetric transformations
$\delta S_{\rm tot}=0$.
The theta parameter appears 
in $T$ and $\BT$ and 
the associated term 
$-{i\eta\over\sqrt{2}}\int_{\partial\Sigma}dx^0 (T\sigma -\BT \Bsigma)$
is rewritten as
$$\eqalign{
&-{i\eta\over\sqrt{2}}\int_{\partial \Sigma}dx^0 (T\sigma -\BT\Bsigma)
={\theta\over2\pi}\int_{\partial \Sigma}dx^0
{\eta\over\sqrt{2}}(\sigma +\Bsigma)+\cdots .
}$$
This boundary term reflects the effect of the theta parameter $\theta$. 
That is consistent with the above consideration.

\newsec{Some extensions}
In this section, we discuss two extensions of previous models on 
supermanifolds. 
The one is the non-abelian extension of the gauge group 
and the other is the $(0,2)$ supersymmetric model. 

\subsec{the model with non-abelian gauge symmetry}

We consider an $\CN = 2$ supersymmetric non-abelian gauge theory 
in two dimensions. 
The supersymmetry is generated 
by supercharges $Q_{\pm}$ and $\BQ_{\pm}$ defined by \Scharge.
In order to construct the non-abelian gauge theory, we introduce 
a vector superfield $V$ which consists of 
gauge multiplets. 
It is a Lie algebra valued superfield and can be expressed by component fields 
in Wess-Zumino gauge in the same fashion of \vects.
Here the lowest components $v_i$ $(i=0,1)$ are vector potentials 
of non-abelian gauge field 
and its field strength is defined as 
$v_{01}=\partial_0 v_1-\partial_1v_0+i[v_0,v_1]$.
Using this vector superfield and the super derivatives \Sder, 
we can define covariant derivatives 
$\CD_\pm$ and $\BCD_\pm$ as
$$
\CD_\pm = e^{-V} D_\pm e^{+V},\quad 
\BCD_\pm =e^{+V} \BD_\pm e^{-V}.
$$
$\CD_-$ and $\BCD_+$ are described concretely in component expansion as 
$$\eqalign{
\CD_- &= {\partial \over \partial \theta^-} - i\Btheta^-\partial_- +\CV_-,\quad 
\BCD_+ = -{\partial \over \partial\Btheta_+}+ i\theta^+\partial_+ +\BCV_+,\cr
\CV_- &= \Btheta^- v_{-} - \sqrt{2}\Btheta^+\sigma 
+2i\theta^+\Btheta^-\Blambda_-
+2i\theta^+\Btheta^+\Blambda_+ -2i\Btheta^-\Btheta^+\lambda_- \cr
&\phantom{=} -\theta^+\Btheta^-\Btheta^+ 
\biggl(2D-i\partial_{-}v_{+}+ {1 \over 2}[v_-,v_+]
-[\sigma ,\Bsigma]\biggr)\cr
&\phantom{=} -i\sqrt{2}\theta^- \Btheta^- \Btheta^+ (D_0-D_1)\sigma 
- 2\theta^+\theta^-\Btheta^-\Btheta^+ 
\bigl((D_0-D_1)\Blambda_+ +i\sqrt{2}[\sigma ,\Blambda_-]\bigr),\cr
\BCV_+ &= -\theta^+ v_{+}+\sqrt{2}\theta^-\sigma 
-2i\theta^+\theta^-\Blambda_+
-2i\theta^+\Btheta^- {\lambda}_{+}-2i\theta^-\Btheta^-\lambda_{-}\cr
&\phantom{=} +\theta^+\theta^-\Btheta^- 
\biggl(-2D-i\partial_{+}v_{-}+{1 \over 2}[v_{+},v_{-}]
-[\sigma ,\Bsigma]\biggr)\cr
&\phantom{=} +i\sqrt{2}\theta^+ \theta^- \Btheta^+ (D_0+D_1)\sigma 
-2\theta^+\theta^-\Btheta^-\Btheta^+ 
\bigl((D_0+D_1)\lambda_- + i\sqrt{2}[\sigma ,\lambda_+]\bigr),
}$$
where $\partial_\pm = \partial_0 \pm \partial_1$ and 
$v_{\pm}=v_0 \pm v_1$. $D_j$ $(j=0,1)$ is an ordinary gauge covariant derivative.
Then the gauge invariant field strength $\Sigma$ is constructed 
by $\CD_-$ and $\BCD_+$, so that in component expansion
$$\eqalign{
\Sigma &= {1 \over 2\sqrt{2}}\{\BCD_+,\CD_-\}\cr
&=\sigma +i\sqrt{2}\theta^+\Blambda_+-i\sqrt{2}\Btheta^- \lambda_- 
+\sqrt{2}\theta^+\Btheta^- (D-iv_{01})\cr
&\phantom{=} +i\theta^-\Btheta^- (D_0-D_1)\sigma 
-i\theta^+\Btheta^+ (D_0+D_1)\sigma 
-\sqrt{2}\theta^+\theta^-\Btheta^- 
\bigl((D_0-D_1)\Blambda_+ + i\sqrt{2}[\sigma ,\Blambda_-]\bigr)\cr
&\phantom{=} +\sqrt{2}\theta^+\Btheta^-\Btheta^+ 
\bigl((D_0+D_1)\lambda_{-}+i\sqrt{2}[\sigma ,\lambda_+]\bigr)\cr
&\phantom{=} -\theta^+\theta^-\Btheta^-\Btheta^+ 
\bigl(D^mD_m\sigma +[\sigma ,[\sigma ,\Bsigma]]+i[\partial^mv_m ,\sigma]\bigr).
}$$
It is a twisted chiral superfield 
and the Lagrangian for this gauge multiplet is denoted by 
\eqn\NALg{\eqalign{
L_{\rm gauge} &=-{1 \over e^2}\int d^4\theta\ \Tr\BSigma\Sigma\cr
&= {1 \over e^2} \Tr\biggl(
-D_i\Bsigma D^i\sigma 
-{1 \over 2}[\sigma ,\Bsigma]^2+ {1 \over 2}(D^2+v_{01}^2)\cr
&\phantom{=} +{i \over 2}(\Blambda_+ D_{-}\lambda_+ 
- D_- \Blambda_+ \cdot \lambda_{+} +\Blambda_- D_+ \lambda_- 
-D_+\Blambda_- \cdot \lambda_-)\cr
&\phantom{=} +\sqrt{2}[\sigma ,\Blambda_-]\lambda_+
-\sqrt{2}\Blambda_+[\Bsigma, \lambda_-]\biggr).
}}
One more possible term 
is the FI term $L_{D,\theta}$, 
\eqn\NALdt{\eqalign{
L_{D,\theta}&={it \over 2\sqrt{2}}
\int d\theta^+d\Btheta^-\Tr\Sigma \bigg|_{\theta^-=\Btheta^+=0}
-{i {\bar t} \over 2\sqrt{2}}
\int d\theta^-d\Btheta^+\Tr\BSigma\bigg|_{\theta^+=\Btheta^-=0} \cr
&= \Tr\biggl(-rD+{\theta \over 2\pi}v_{01}\biggr).
}}
Here we pick up the trace part of $\Sigma$ in constructing $L_{D,\theta}$ 
and consider only diagonal $U(1)$ part in $L_{D,\theta}$.
This term is gauge invariant and characterised 
by the theta parameter $\theta$ and 
the FI parameter $r$ of this diagonal $U(1)$. 

The supersymmetric transformations for the vector superfield are 
represented in component fields
$$\eqalign{
\delta \sigma &=-i\sqrt{2}(\epsilon_-\Blambda_++\Bepsilon_+\lambda_-),\quad
\delta \Bsigma =-i\sqrt{2}(\epsilon_+\Blambda_-+\Bepsilon_-\lambda_+),\cr
\delta (v_0-v_1) &=2i (\epsilon_-\Blambda_-+\Bepsilon_-\lambda_-),\quad 
\delta (v_0+v_1) =2i (\epsilon_+\Blambda_++\Bepsilon_+\lambda_+),\cr
\delta D &=\epsilon_+((D_0-D_1)\Blambda_++i\sqrt{2}[\sigma ,\Blambda_-])
+\epsilon_-((D_0+D_1)\Blambda_-+i\sqrt{2}[\Bsigma,\Blambda_+])\cr
&\phantom{=} -\Bepsilon_+((D_0-D_1)\lambda_++i\sqrt{2}[\Bsigma ,\lambda_-])
-\Bepsilon_-((D_0+D_1)\lambda_-+i\sqrt{2}[\sigma ,\lambda_+]),\cr
\delta {\lambda}_+ &=i\epsilon_+(D+iv_{01}+[\sigma ,\Bsigma])
+\sqrt{2}\epsilon_-(D_0+D_1)\Bsigma,\cr
\delta {\lambda}_- &=i\epsilon_-(D-iv_{01}-[\sigma ,\Bsigma])
+\sqrt{2}\epsilon_+(D_0-D_1)\sigma,\cr
\delta \Blambda_+ &=-i\Bepsilon_+(D-iv_{01}+[\sigma ,\Bsigma])
+\sqrt{2}\Bepsilon_-(D_0+D_1)\sigma,\cr
\delta \Blambda_- &=-i\Bepsilon_-(D+iv_{01}-[\sigma ,\Bsigma])
+\sqrt{2}\Bepsilon_+(D_0-D_1)\Bsigma.
}$$

Next we consider matter parts of this theory. 
We here use the bosonic chiral and anti-chiral superfields \bcs\ and \bacs, 
and their fermionic counter parts \fcs\ and \facs. 
The kinetic terms for these chiral superfields are evaluated as
\eqn\NALb{\eqalign{
L_{\rm kin}^b &= \int d^2\theta d^2\Btheta
\sum_I\BPhi^Ie^{2 V}\Phi^I\cr
&= \sum_I\biggl[-(\BD_j\Bphi^I)(D^j\phi^I) +\BF^IF^I 
-\Bphi^I\{\sigma, \Bsigma\}\phi^I 
+\Bphi^I  D\phi^I \cr
&\phantom{=} +{i \over 2}
\bigl(\Bpsi^I_+D_-\psi^I_+ -(\BD_-\Bpsi^I_+) \psi^I_+\bigr)
+{i \over 2}\bigl(\Bpsi^I_-D_+\psi^I_- -(\BD_+\Bpsi^I_-) \psi^I_-\bigr)\cr
&\phantom{=}+ i\sqrt{2}\Bphi^I(\lambda_+\psi^I_- 
	-\lambda_-\psi^I_+)
-i\sqrt{2}(\Bpsi^I_- \Blambda_ +-\Bpsi^I_+ \Blambda_-)\phi^I \cr
&\phantom{=}-\sqrt{2}(\Bpsi^I_- \sigma \psi^I_+ 
	+\Bpsi^I_- \Bsigma \psi^I_-)\biggr]
}}
for the bosonic chiral superfields.
In the same way, we obtain the kinetic term 
of the fermionic chiral superfields, 
\eqn\NALf{\eqalign{
L_{\rm kin}^f &= \int d^2\theta d^2\Btheta \sum_A \BXi^Ae^{2 V}\Xi^A\cr
&= \sum_A \biggl[
-(\BD_j\Bxi^A)(D^j\xi^A)+\Bchi^A\chi^A 
	-\Bxi^A\{\sigma, \Bsigma\}\xi^A 
+\Bxi^A D \xi^A \cr
&\phantom{=} +{i \over 2}\bigl(\Bb^A_+D_-b^A_+ -(\BD_-\Bb^A_+) b^A_+\bigr)
+{i \over 2}\bigl(\Bb^A_-D_+b^A_- -(\BD_+\Bb^A_-) b^A_-\bigr) \cr
&\phantom{=}+i\sqrt{2}\Bxi^A(\lambda_+b^A_- -\lambda_-b^A_+)
-i\sqrt{2}(\Bb^A_-\Blambda_+-\Bb^A_+\Blambda_-)\xi^A \cr
&\phantom{=}-\sqrt{2}(\Bb^A_- \sigma b^A_+ 
	+\Bb^A_+ \Bsigma b^A_-)\biggr] .
}}
With the terms described in \NALg, \NALdt, \NALb\ and \NALf, 
we can construct the action of this theory
$$
L = L_{\rm gauge} + L_{D,\theta} + L_{\rm kin}^b + L_{\rm kin}^f.
$$
It has $\CN=2$ supersymmetry generated by \Scharge.

Now let us consider a concrete example of this non-abelian model.
We set that the gauge group 
$G= U(N)$ and take the bosonic chiral superfields $\Phi^I{}^k$ 
and the fermionic chiral superfields 
$\Xi^A{}^k$ $(k=1,2,\cdots ,N; I=1,2,\cdots ,m; A=1,2,\cdots ,n)$. Here the 
index $k$ labels the $N$ dimensional representation of $G$.
Also the global symmetry labelled by indices $I$ and $A$ 
commute with the action of $G$. 
The potential of this model is determined by $L_U$, 
$$\eqalign{
L_U &= {1 \over 2e^2}\Tr D^2-r\Tr D+
\sum_I \Bphi^I  D \phi^I+\sum_A \Bxi^A  D \xi^A\cr
&\phantom{=} -{1 \over 2e^2}\Tr [\sigma ,\Bsigma]^2
-\sum_I\Bphi^I\{\sigma, \Bsigma\}\phi^I
-\sum_A \Bxi^A\{\sigma, \Bsigma\}\xi^A .
}$$
By eliminating the D-field, the potential $U$ can be read
\eqn\NApot{\eqalign{
U &={e^2 \over 2}
\sum_{k,l=1}^N \biggl[\sum_I {\Bphi^I}_k  \phi^I{}^l
+\sum_A  {\Bxi^A}_k \xi^A{}^l 
-r\delta_k{}^l \biggr] \biggl[\sum_I {\Bphi^I}_l  \phi^I{}^k
+\sum_A  {\Bxi^A}_l \xi^A{}^k 
-r\delta_l{}^k \biggr]\cr
&\phantom{=} +{1 \over 2e^2}\Tr [\sigma ,\Bsigma]^2 
	+\sum_I \Bphi^I\{\sigma, \Bsigma\}\phi^I
+\sum_A \Bxi^A\{\sigma, \Bsigma\}\xi^A.
}}
The space of classical vacua is determined by a vanishing locus of $U$ 
up to gauge transformation. 
The action of $G$ can be regarded as the transformation 
of the direct product $Y$ of $N$ $\Bbb{C}^{m|n}$'s. 
The first term in the potential \NApot\  is the square 
of the moment map for the action of $U(N)$ 
on $Y$.  From the constraint $U=0$, $\sigma$ must vanish 
to describe the classical vacua 
and $\phi^I{}^k$ and $\xi^A{}^k$ must satisfy constraints
\eqn\naDflat{
- {1 \over e^2}D^l{}_k=\sum_{I}(\Bphi^I)_k(\phi^I)^l
	+\sum_A (\Bxi^A)_k (\xi^A)^l
	-r\delta_k{}^l =0, \quad k,l=1,2,\cdots ,N.
}
It means that the vectors in $\Bbb{C}^{m|n}$ represented by the rows 
$(\phi^I{}^k,\xi^A{}^k)$ $(k=1,2,\cdots ,N)$ 
divided by $\sqrt{r}$ are orthonormal. 
That is to say, the space of classical vacua is the  
gauge invariant subspace of $Y$ characterised 
by above orthonormal vectors.
When we restrict ourselves to the bosonic sector by putting $\xi^A{}^k=0$, 
this subspace (classical vacua) turns out to be the Grassmannian Gr($N,m$).

\subsec{quantum property of the non-abelian model}

The classical vacua are determined by the D-flatness condition \naDflat\ 
for $r\gg 0$. It requires that the fields $\phi^I$ and  $\xi^A$ have 
vacuum expectation values. 
We discuss the quantum correction 
for D-field at the one-loop level. From the quantum effects, the operator 
${\cal O}^l{}_k=\sum_I\Bphi^I{}_k \phi^I{}^l + \sum_A\Bxi^A{}_k \xi^A{}^l$ 
can have an expectation value. 
If it is equal to $r\delta_k{}^l$, 
$D^l{}_k$ can vanish even though 
$\phi^I$ and $\xi^A$ do not have any expectation values.
In such a case, one could obtain vacua from $U=0$. 
We shall study this situation here. 

We shall evaluate the 
expectation values of $D^l{}_k$ 
by performing the loop calculation about 
$\phi^I$ and $\xi^A$. In free field approximation, 
the mass terms for $\phi^I$ and $\xi^A$ are written as 
$$\eqalign{
\sum_I  \Bphi^I\{\sigma, \Bsigma\}\phi^I
+\sum_A  \Bxi^A\{\sigma, \Bsigma\}\xi^A.
}$$
In terms of this approximation 
for $\phi$ and $\xi$, we can evaluate the 
expectation value of $\CO$ with a cut-off parameter $\mu$, 
$$\eqalign{
\langle \CO \rangle_{\hbox{\sevenrm 1-loop}} &=
\sum_{I=1}^n\int {d^2k \over (2\pi)^2}{1 \over k^2+\{\sigma ,\Bsigma\}}
-\sum_{A=1}^m \int {d^2k \over (2\pi)^2}{1 \over k^2+\{\sigma ,\Bsigma\}}\cr
&=-{1 \over 4\pi}(m-n)
\log \biggl({\{\sigma ,\Bsigma\} \over 2\mu^2}\biggr) . 
}$$
If $m-n$ is equal to zero, the quantum effect vanishes 
at least in this approximation.
For $m-n \neq 0$, we have to analyse 
the D-flatness condition expressed as 
$$\eqalign{
-{1\over {4\pi}}(m-n)\log\biggl({ \{\sigma ,\Bsigma\}\over {2\mu^2}}\biggr)
-r=0 ,
}$$
or equivalently 
\eqn\NAvacua{
\{\sigma ,\Bsigma\}=2\mu^2\exp\biggl(-{4\pi r \over m-n }\biggr).
}
In order to discuss vacua, we have to consider another condition 
$\Tr [\sigma ,\Bsigma]^2=0$ from the potential $U=0$. 
Here we may put $[\sigma ,\Bsigma]=0$ to describe the vacua 
and rewrite \NAvacua\ as
$$
\sigma\Bsigma=\mu^2\exp\biggl({-4\pi r \over {m-n}}\biggr) .
$$
It means that $\sigma = \sigma_0 \mu\exp (-{2\pi r / (m-n)})$ 
with a unitary matrix $\sigma_0$. 
This expectation value of $\sigma$ gives a positive mass squared 
to the $\phi^I$ and $\xi^A$ 
and these fields have zero expectation values.

\subsec{$(0,2)$ model}

In this subsection, we consider $(0,2)$ supersymmetric model with abelian gauge symmetry 
in two dimensions. 
The supersymmetry is generated 
by two supercharges $Q_+$, $\BQ_+$ in the right-moving sector. 
They commute with super derivatives $D_+$, $\BD_+$ and 
associated fermionic coordinates in a world-sheet superspace are given 
by $\theta^+$ and $\Btheta^+$.
In order to formulate a gauge theory, we introduce gauge fields 
$v_i$ $(i=0,1)$. 
It is described by a superfield $\Psi$ which  
can be written in the Wess-Zumino gauge as 
$\Psi =\BPsi= \theta \Btheta (v_0+v_1)$. 
With this $\Psi$, gauge covariant derivatives are denoted by 
$$\eqalign{
&\CD_+=e^{-\Psi} D_+ e^{+\Psi}
= {\partial \over \partial\theta^+} 
	-i \Btheta^+ (D_0 + D_1),\cr
&\BCD_+=e^{+\BPsi} \BD_+ e^{-\BPsi}
= - {\partial \over \partial \Btheta^+} 
	+ i\theta^+ (D_0 + D_1),\cr
&\CD_0 - \CD_1 = \partial_0 - \partial_1 + i(v_0-v_1)
+2\theta^+ \Blambda_- + 2\Btheta^+\lambda_- + 2i\theta^+\Btheta^+ D.
}$$
The gauge invariant field strength $\Upsilon$ is defined as 
$$\eqalign{
\Upsilon =[\BCD_{+}, \CD_0 - \CD_1] =
	-2\lambda_- + 2i\theta^+ (D - iv_{01})
	+2i\theta^+ \Btheta^+ \partial_+ \lambda_- ,
}$$
and the kinetic term of this gauge multiplet is 
\eqn\oIIgauge{
L_{\rm gauge} = {1 \over 8e^2} 
\int d\theta^+ d\Btheta^+ \Upsilon \BUpsilon
= {1 \over e^2}\biggl({1 \over 2}D^2 + {1 \over 2} v_{01}^2 
+ i\Blambda_-(\partial_0 + \partial_1)\lambda_- \biggr).
}
One more possible term of abelian gauge theory is the FI term, 
\eqn\oIIdt{
L_{D,\theta} = {t \over 4} \int d\theta^+ 
\Upsilon \bigg|_{\Btheta^+=0}
+ {{\bar t} \over 4}\int d\Btheta^+ \BUpsilon\bigg|_{\theta^+=0}
=-rD+{\theta \over 2\pi}v_{01},
}
with the FI parameter $t=ir + \theta / (2\pi)$.

We also introduce matter multiplets. 
Bosonic chiral superfield $\Phi^I$ and its conjugate $\BPhi^I$ are 
defined in component expansion as  
$$\eqalign{
\Phi^I &=\phi^I +\sqrt{2}\theta^+\psi^I_+ -i\theta^+\Btheta^+ (D_0+D_1)\phi^I , \cr
\BPhi^I &= \Bphi^I - \sqrt{2}\Btheta^+\Bpsi^I_+ 
	+i\theta^+\Btheta^+ (D_0+D_1)\Bphi^I .
}$$
In our case there are fermionic chiral and anti-chiral superfields, 
$$\eqalign{
\Xi^A &= \xi^A +\sqrt{2}\theta^+ b^A_+ -i\theta^+\Btheta^+ (D_0+D_1)\xi^A ,\cr
\BXi^A &= \Bxi^A + \sqrt{2}\Btheta^+\Bb^A_+ 
	+i\theta^+\Btheta^+ (D_0+D_1)\Bxi^A . \cr
}$$
The kinetic term $L_{\rm kin}$ of these fields is defined as
\eqnn\oIIkin
$$\eqalignno{
L_{\rm kin} &= L_{\rm kin}^b + L_{\rm kin}^f , &\oIIkin \cr
L_{\rm kin}^b &= {i \over 2}\int d\theta^+ d\Btheta^+ 
	\sum_I\BPhi^I (\CD_0 - \CD_1)\Phi^I \cr
&= \sum_I \Bigl[- D_j \Bphi^I D^j\phi^I + i\Bpsi^I_+ (D_0-D_1) \psi^I_+ 
	+Q_I D\Bphi^I\phi^I \cr
&\phantom{=} -i\sqrt{2}Q_I \Bphi^I\lambda_- \psi^I_+
	+i\sqrt{2}Q_I\Bpsi^I_+ \Blambda_- \phi^I \Bigr] , \cr
L_{\rm kin}^f &= {i \over 2}\int d\theta^+ d\Btheta^+ 
	\sum_A \BXi^A(\CD_0 - \CD_1)\Xi^A \cr
&=\sum_A \Bigl[- D_j \Bxi^A D^j\xi^A +i\Bb^A_+ (D_0-D_1) b^A_+ 
         +q_A D\Bxi^A\xi^A \cr
&\phantom{=} -i\sqrt{2}q_A \Bxi^A \lambda_- b^A_+
	+i\sqrt{2} q_A \Bb^A_+ \Blambda_- \xi^A \Bigr] .
}$$
We take other types of matter multiplet $\Lambda_{-a}$ and 
${\tilde \Lambda}_{-{\tilde a}}$, where $\Lambda_{-a}$'s are fermionic superfields and 
${\tilde \Lambda}_{-{\tilde a}}$'s are bosonic ones.  
They are expanded in terms of $\theta^+$ and $\Btheta^+$, 
$$\eqalign{
\Lambda_{-a} &= \lambda_{-a} -\sqrt{2}\theta^+ G_a 
	-i\theta^+\Btheta^+ (D_0+D_1)\lambda_{-a}
	-\sqrt{2}\Btheta^+ E_a(\Phi, \Xi) , \cr
{\tilde \Lambda}_{-{\tilde a}} &= {\tilde \lambda}_{-{\tilde a}} 
-\sqrt{2}\theta^+ {\tilde G}_{\tilde a} -
	i\theta^+\Btheta^+ (D_0 + D_1) {\tilde \lambda}_{-{\tilde a}}
	-\sqrt{2}\Btheta^+ {\tilde E}_{\tilde a} (\Phi, \Xi) ,\cr
E_a(\Phi, \Xi) &= E_a(\phi, \xi) 
	+\sqrt{2}\theta^+ \sum_I\psi^I_+ {\partial E_a \over \partial \phi^I}
	+\sqrt{2}\theta^+ \sum_A b^A_+ {\partial E_a \over \partial \xi^A}
	-i\theta^+\Btheta^+ (D_0 + D_1)E_a ,\cr
{\tilde E}_{\tilde a}(\Phi, \Xi) &= {\tilde E}_{\tilde a} (\phi, \xi) 
	+\sqrt{2}\theta^+ \sum_I \psi^I_+ 
	{\partial {\tilde E}_{\tilde a} \over \partial \phi^I}
	+\sqrt{2}\theta^+ \sum_A b^A_+ 
	{\partial {\tilde E}_{\tilde a} \over \partial \xi^A}
	-i\theta^+\Btheta^+ (D_0 + D_1){\tilde E}_{\tilde a} ,
}$$
where $E_a(\Phi,\Xi)$ and ${\tilde E}_{\tilde a}(\Phi,\Xi)$ are superfields.
The kinetic term of these fields is defined as
\eqn\oIIEkin{\eqalign{
L_{\Lambda} &= {1 \over 2}\int d\theta^+ d\Btheta^+ 
	(\BLambda_{-a}\Lambda_{-a} +\overline{\tilde \Lambda}_{-\tilde a }
	{\tilde \Lambda}_{-\tilde a }) \cr
&= {1 \over 2}i\Blambda_{-a}(D_0 + D_1)\lambda_{-a}
	-{1 \over 2}i\bigl((D_0 + D_1)\Blambda_{-a}\bigr)\lambda_{-a}
	+\BG_a G_a - \BE_a E_a 	\cr
&\phantom{=}
	+{1 \over 2}i\overline{\tilde \lambda}_{-\tilde a}
	(D_0+D_1){\tilde \lambda}_{-\tilde a }
	-{1 \over 2}i\bigl((D_0+D_1)\overline{\tilde \lambda}_{-\tilde a}\bigr)
	{\tilde \lambda}_{-\tilde a} 
	+\overline{\tilde G}_{\tilde a} {\tilde G}_{\tilde a}
	-\overline{\tilde E}_{\tilde a} {\tilde E}_{\tilde a} \cr
&\phantom{=}-\sum_I \biggl[ 
	\Blambda_{-a}\psi^I_+ {\partial E_a \over \partial \phi^I}
	+\overline{\partial E_a \over \partial \phi^I}\Bpsi^I_+ \lambda_{-a}
	+\overline{\tilde \lambda}_{-\tilde a}\psi^I_+ 
	{\partial {\tilde E}_{\tilde a} \over \partial \phi^I}
	+\overline{\partial {\tilde E}_{\tilde a} \over \partial \phi^I} \Bpsi^I_+ 
	{\tilde \lambda}_{-\tilde a} \biggr] \cr
&\phantom{=}-\sum_A \biggl[ 
	\Blambda_{-a}b^A_+ {\partial E_a \over \partial \xi^A} 
	+\overline{\partial E_a \over \partial \xi^A}\Bb^A_+ \lambda_{-a} 
	+\overline{\tilde \lambda}_{-\tilde a} b^A_{+} 
	{\partial {\tilde E}_{\tilde a} \over \partial \xi^A}
	+\overline{\partial {\tilde E}_{\tilde a} \over \partial \xi^A} 
	\Bb^A_+ {\tilde \lambda}_{-\tilde a} \biggr] .
}}
Next we introduce superfields 
$J^a(\Phi, \Xi)$ and ${\tilde J}^{\tilde a}(\Phi, \Xi)$ with relations, 
\eqn\EJrel{
E_a J^a + {\tilde E}_{\tilde a} {\tilde J}^{\tilde a}=0,\quad 
\BCD_+ J^a = 0 ,\quad \BCD_+ {\tilde J}^{\tilde a} = 0 . 
}
$J^a$'s are bosonic fields and ${\tilde J}^{\tilde a}$'s are fermionic ones. 
Since the superfields $\Lambda_{-a}$ and ${\tilde \Lambda}_{-{\tilde a}}$ 
satisfy the relations,
$$
\BCD_+ \Lambda_{-a} = \sqrt{2} E_a(\Phi, \Xi) ,\quad 
\BCD_+ {\tilde \Lambda}_{-{\tilde a}} = \sqrt{2} {\tilde E}_{\tilde a}(\Phi, \Xi) ,
$$
which lead to 
$\BCD_+ (\Lambda_{-a} J^a + {\tilde \Lambda}_{-{\tilde a}} {\tilde J}^{\tilde a})=0$, 
we can construct another term $L_{J}$, 
\eqn\oIIEJ{\eqalign{
L_J &= {1 \over \sqrt{2}}\int d\theta^+ (\Lambda_{-a}J^a
	+{\tilde \Lambda}_{-{\tilde a}}{\tilde J}^{\tilde a})\bigg|_{\Btheta^+ = 0}
	+{1 \over \sqrt{2}}\int d\Btheta^+ (\BJ^a \BLambda_{-a}
	+\overline{\tilde J}^{\tilde a} \overline{\tilde \Lambda}_{-{\tilde a}})
	\bigg|_{\theta^+ = 0} \cr
&= \sum_I \biggl[\psi^I_+ \lambda_{-a} {\partial J^a \over \partial \phi^I}
	+\psi^I_+{\tilde \lambda}_{-\tilde a} {\partial {\tilde J}^{\tilde a} \over \partial \phi^I}
	+\overline{\partial J^a \over \partial \phi^I} \Blambda_{-a} \Bpsi^I_+ 
	+\overline{\partial {\tilde J}^{\tilde a} \over \partial \phi^I} 
	\overline{\tilde \lambda}_{-\tilde a} \Bpsi^I_+ \biggr] \cr
&\phantom{=} +\sum_A\biggl[
	-b^A_+\lambda_{-a} {\partial J^a \over \partial \xi^A} 
	+b^A_+{\tilde \lambda}_{-\tilde a} {\partial {\tilde J}^{\tilde a} \over \partial \xi^A} 
	-\overline{\partial J^a \over \partial \xi^A} \Blambda_{-a} \Bb^A_+ 
	+\overline{\partial {\tilde J}^{\tilde a} \over 
	\partial \xi^A} \overline{\tilde \lambda}_{-\tilde a} \Bb^A_+ \biggr]\cr
&\phantom{=}-G_aJ^a  -\BJ^a\BG_a 
	-{\tilde G}_{\tilde a}{\tilde J}^{\tilde a} 
	-\overline{\tilde J}^{\tilde a} \overline{\tilde G}_{\tilde a} .
}}
Here we assume $E_a$, ${\tilde E}_{\tilde a}$, $J^a$ and ${\tilde J}^{\tilde a}$
are holomorphic functions of $\Phi^I$ and $\Xi^A$. 
By collecting the terms written 
in \oIIgauge, \oIIdt, \oIIkin, \oIIEkin\ and \oIIEJ, 
we obtain the total Lagrangian of this model, 
$$
L=L_{\rm gauge}+L_{\rm kin}+L_{\Lambda}+L_{D,\theta}+L_J .
$$

We now investigate the reduction of $(2,2)$ multiplets into 
$(0,2)$ multiplets. 
Firstly the gauge multiplet $\Sigma$ of $(2,2)$ model is reduced to 
$\Sigma'$ by putting $\theta^- = \Btheta^- = 0$, 
$$
\Sigma' = \Sigma \big|_{\theta^- = \Btheta^- = 0}.
$$
Next the $(2,2)$ chiral superfields $\Phi^I$ and $\Xi^A$ 
are respectively reduced to 
$(0,2)$ multiplets $\Phi'_I$ and $\Xi'_A$, 
$$\eqalign{
\Phi'_I &= \Phi^I \big|_{\theta^-=\Btheta^- =0},\quad 
\Xi'_A =\Xi^A\big|_{\theta^-=\Btheta^- =0}.\cr
}$$
By using these fields, we can introduce $\Lambda_{-I}$ and 
${\tilde \Lambda}_{-A}$ as
\eqn\leloii{\eqalign{
\Lambda_{-I} &= {1 \over \sqrt{2}}\CD_- 
	\Phi^I \bigg|_{\theta^- =\Btheta^- =0},\quad 
{\tilde \Lambda}_{-A} ={1 \over \sqrt{2}}
\CD_-\Xi^A \bigg|_{\theta^- =\Btheta^- =0} .
}}
Then we have the relations, 
\eqn\reloii{\eqalign{
\BCD_+\Lambda_{-I} = \sqrt{2}E_I, &\quad 
E_I=\sqrt{2}Q_I\Sigma'\Phi'_I, \cr
\BCD_+ {\tilde \Lambda}_{-A} = \sqrt{2}{\tilde E}_A, &\quad
{\tilde E}_A = \sqrt{2}q_A\Sigma' \Xi'_A . 
}}
In order to compare this $(0,2)$ model to $(2,2)$ case, 
we take a $(2,2)$ model with a superpotential $W(\Phi ,\Xi)$ 
which is the function of the chiral superfields $\Phi^I$ and $\Xi^A$. 
The chiral superfields $\Phi^I$ and $\Xi^A$ are respectively reduced into 
$(\Phi'_I, \Lambda_{-I})$ and $(\Xi'_A, {\tilde \Lambda}_{-A})$ in 
the $(0,2)$ model. 
They satisfy the relations \leloii\ and \reloii.
The interaction terms in the $(2,2)$ model are determined by 
the superpotential $W$. 
In the $(0,2)$ model, one must specify functions $J^I$ and ${\tilde J}^A$ 
obeying 
\eqn\EJrelp{
E_I J^I + {\tilde E}_A {\tilde J}^A=0.
}
Comparing superpotential terms $L_W$ with $L_J$ in the $(0,2)$ model, 
we can read the correspondence, 
$$\eqalign{
(\lambda_{-I},J^I,G_I) &\leftrightarrow 
\biggl(\psi^I_-, {\partial W \over \partial \phi^I} ,F^I\biggr) ,\cr
({\tilde \lambda}_{-A}, {\tilde J}^A, {\tilde G}_A)
&\leftrightarrow \biggl(b^A_-, {\partial W \over \partial \xi^A}, \chi^A\biggr) .
}$$
Under this situation, the relation \EJrelp\ is
equivalent to quasi-homogeneous condition, 
$$
\sum_I Q_I \phi^I {\partial W \over \partial \phi^I} 
+ \sum_A q_A \xi^A {\partial W \over \partial \xi^A} = 0 .
$$
It means that 
$W(\lambda^{Q_I}\phi^I ,\lambda^{q_A}\xi^A)=W(\phi^I,\xi^A)$ 
$(\forall \lambda \in \Bbb{C}^\times)$ for the superpotential $W(\Phi^I ,\Xi^A)$. 

Next we write down the potential terms of this model 
containing auxiliary fields $D$, 
$G_I$ and ${\tilde G}_A$, 
$$\eqalign{
L_U &= {1 \over 2e^2} D^2 -rD +\sum_I Q_I D \Bphi^I \phi^I
	+\sum_A q_A D \Bxi^A \xi^A \cr
&\phantom{=} + \sum_A \bigl(\BG_I G_I -\BE_I E_I -G_I J^I -\BJ^I \BG_I \bigr)
	+ \sum_A \bigl(\overline{\tilde G}_A {\tilde G}_A
	-\overline{\tilde E}_A {\tilde E}_A
	 -{\tilde G}_A{\tilde J}^A
	-\overline{\tilde J}^A \overline{\tilde G}_A \bigr) .\cr
}$$
After eliminating these auxiliary fields $D$, $G_I$ and ${\tilde G}_A$, 
the potential term is expressed as 
$$
U= {e^2 \over 2}\biggl(
	\sum_I Q_I \Bphi^I\phi^I + \sum_A q_A \Bxi^A\xi^A -r\biggr)^2
	+\BE_I E_I +\overline{\tilde E}_A{\tilde E}_A +J^I\BJ^I
	+{\tilde J}^A\overline{\tilde J}^A.
$$
If we put $J^I=\partial W / \partial \phi^I$ and 
${\tilde J}^A = \partial W / \partial \xi^A$ into the above formula, 
the potential $U$ can be written as \potII.
One can evaluate the classical vacua by analysing this potential $U$. 

Now let us return to the $(0,2)$ model and 
study the spin $1/2$ part. 
The scalar part (spin $0$ part) of the 
 $(0,2)$ model is dominated at the low energy 
by the condition $U=0$, which determines the structure of vacua.  
In addition, there are spin $1/2$ fields in supersymmetric model and 
massless fields play an important role in the dynamics of the 
low energy physics. 
We shall concentrate on the massless fields in this model. 

The mass terms of fields with spin $1/2$ at the low energy are induced by 
the Yukawa couplings. 
The related parts in the Lagrangian are described 
by $E_a$, ${\tilde E}_{\tilde a}$, $J^a$ and  ${\tilde J}^{\tilde a}$, 
$$\eqalign{
L &=-\sqrt{2}i\lambda_- \biggl( \sum_I Q_I \psi^I_+ \Bphi^I
		- \sum_A q_A b^A_+ \Bxi^A \biggr)
+\sqrt{2}i \biggr(\sum_I Q_I \Bpsi^I_+ \phi^I
		-\sum_A q_A \Bb^A_+\xi^A \biggr) \Blambda_- \cr
&\phantom{=}+ \pmatrix{\psi^I_+ & b^A_+}
\pmatrix{\displaystyle {\partial E_a\over \partial \phi^I} & 
	\displaystyle -{\partial{\tilde E}_{\tilde a}\over \partial \phi^I}\cr
	\displaystyle {\partial E_a\over \partial \xi^A} & 
	\displaystyle -{\partial {\tilde E}_{\tilde a}\over \partial \xi^A}}
\pmatrix{\Blambda_{-a} \cr \overline{\tilde \lambda }_{-\tilde a}}
+\pmatrix{\psi^I_+ & b^A_+}
\pmatrix{\displaystyle {\partial J^a \over \partial \phi^I} &
	\displaystyle  {\partial{\tilde J}^{\tilde a}\over \partial \phi^I}\cr
	\displaystyle {\partial J^a\over \partial \xi^A} &
	\displaystyle  {\partial {\tilde J}^{\tilde a}\over \partial \xi^A}}
\pmatrix{\lambda_{-a} \cr {\tilde \lambda }_{-\tilde a}} \cr
&\phantom{=}+\pmatrix{\lambda_{-a} & {\tilde \lambda}_{-\tilde a}}
\pmatrix{\displaystyle \overline{\partial E_a\over \partial \phi^I} & 
	\displaystyle \overline{\partial E_{a}\over \partial \xi^A}\cr
	\displaystyle -\overline{\partial{\tilde E}_{\tilde a} \over \partial \phi^I} & 
	\displaystyle  -\overline{\partial {\tilde E}_{\tilde a}\over \partial \xi^A}}
\pmatrix{\Bpsi_+^I \cr \Bb^A_+}
+\pmatrix{\Blambda_{-a} & \overline{\tilde \lambda}_{-\tilde a}} 
\pmatrix{
	\displaystyle \overline{\partial J^a\over \partial \phi^I} & 
	\displaystyle \overline{\partial J^a\over \partial \xi^A}\cr
	\displaystyle \overline{\partial{\tilde J}^{\tilde a} \over \partial \phi^I} &
	\displaystyle  \overline{\partial {\tilde J}^{\tilde a}\over \partial \xi^A}}
\pmatrix{\Bpsi_+^I \cr \Bb^A_+ } .
}$$
Here $E_a$ and $J^a$ are Grassmann even, while  
${\tilde E}_{\tilde a}$ and  ${\tilde J}^{\tilde a}$ are Grassmann odd. 
They satisfy the relation 
$E_aJ^a + {\tilde E}_{\tilde a}{\tilde J}^{\tilde a} = 0$ in \EJrel. From this formula, 
the massless right-moving fields $\psi^I_+$ and $b^A_+$ are 
determined by 
$$\eqalign{
\sum_I Q_I \psi^I_+ \Bphi^I - \sum_A q_A b^A_+ \Bxi^A=0 , &\quad
\sum_I Q_I \Bpsi^I_+ \phi^I - \sum_A q_A \Bb^A_+ \xi^A=0 , \cr
\sum_I \psi^I_+ {\partial E_a \over \partial \phi^I}
+\sum_A b^A_+ {\partial E_a \over \partial \xi^A}=0 , &\quad 
\sum_I \psi^I_+ {\partial {\tilde E}_{\tilde a} \over \partial \phi^I}
+\sum_A b^A_+ {\partial {\tilde E}_{\tilde a} \over \partial \xi^A}=0 ,\cr
\sum_I \psi^I_+ {\partial J^a \over \partial \phi^I}
+\sum_A b^A_+ {\partial J^a \over \partial \xi^A}=0 , &\quad 
\sum_I\psi^I_+ {\partial {\tilde J}^{\tilde a} \over \partial \phi^I}
+\sum_A b^A_+ {\partial {\tilde J}^{\tilde a}\over \partial \xi^A}=0 .
}$$
The first two conditions are interpreted as a 
gauge-fixing condition of holomorphic equivalence 
and they define a kind of tangent bundle of $\Bbb{WCP}^{m-1|n}$ 
with $\sum_I Q_I \Bphi^I\phi^I + \sum_A q_A \Bxi^A\xi^A=r$. 
The other conditions mean 
restriction of the bundle to the hypersurfaces 
$E_a=J^a=0$ and ${\tilde E}_{\tilde a}={\tilde J}^{\tilde a}=0$, and 
the right-moving massless fields take their values in this bundle.

Next let us see the left-moving fields 
$\lambda_{-a}$ and ${\tilde \lambda}_{-\tilde a}$. 
Massless conditions of these left-movers are 
expressed by 
$$\eqalign{
\sum_a\lambda_{-a}\overline{\partial E_a\over \partial \phi^I}
-\sum_{\tilde a}{\tilde \lambda}_{-\tilde a}
\overline{\partial {\tilde E}_{\tilde a}\over \partial \phi^I}=0 , &\quad
\sum_a\lambda_{-a}\overline{\partial E_a\over \partial \xi^A}
-\sum_{\tilde a}{\tilde \lambda}_{-\tilde a} \overline{\partial {\tilde E}_{\tilde a}\over \partial \xi^A}
=0 , \cr
\sum_a{\partial J^a\over \partial \phi^I}\lambda_{-a}
+\sum_{\tilde a}
{\partial \tilde{J}^{\tilde a}\over \partial \phi^I}{\tilde \lambda}_{-\tilde a}=0 , &\quad
\sum_a{\partial J^a\over \partial \xi^A}\lambda_{-a}
+\sum_{\tilde a}
{\partial \tilde{J}^{\tilde a}\over \partial \xi^A}{\tilde \lambda}_{-\tilde a}=0 .\cr
}$$
For simplicity, we consider the example in which 
a neutral chiral superfield $\Sigma$ is introduced and the superfields 
$E_a$ and ${\tilde E}_{\tilde a}$ are decomposed as 
$$
E_a=\Sigma S_a(\phi^I,\xi^A) ,\quad 
{\tilde E}_{\tilde a}=\Sigma {\tilde S}_{\tilde a}(\phi^I,\xi^A). 
$$
Then the first relation in \EJrel\ is led to 
$$
S_aJ^a+{\tilde S}_{\tilde a}{\tilde J}^{\tilde a}=0 .
$$
If vacuum expectation value of $\Sigma$ vanishes, 
we do not have to impose the constraints $S_a=0$ and ${\tilde S}_{\tilde a}=0$ 
on the fields $\phi^I$ and $\xi^A$. 
In such a case, the right-moving massless fields $\psi^I_+$ and $b^A_+$ 
are characterised by $J^a={\tilde J}^{\tilde a}=0$ and 
$\sum_IQ_I|\phi^I|^2 + \sum_A q_A \Bxi^A \xi^A=r$. 
On the other hand, the massless conditions for the 
left-movers $\lambda_{-a}$ and ${\tilde \lambda}_{-\tilde a}$ are rewritten as 
\eqna\LSconst
$$\eqalignno{
&\sum_a\lambda_{-a}\overline{S}_a
-\sum_{\tilde a}{\tilde \lambda}_{-\tilde a}
\overline{\tilde S}_{\tilde a}=0 , &\LSconst a \cr
&\sum_a{\partial J^a\over \partial \phi^I}\lambda_{-a}
+\sum_{\tilde a}
{\partial \tilde{J}^{\tilde a}\over \partial \phi^I}{\tilde \lambda}_{-\tilde a}=0 , &\LSconst b \cr
&\sum_a{\partial J^a\over \partial \xi^A}\lambda_{-a}
+\sum_{\tilde a}
{\partial \tilde{J}^{\tilde a}\over \partial \xi^A}{\tilde \lambda}_{-\tilde a}=0 . & \LSconst c \cr
}$$
The constraint \LSconst{a} means a gauge fixing condition for the equivalence 
$(\lambda_{-a}, {\tilde \lambda}_{-\tilde a})
\sim (\lambda_{-a}, {\tilde \lambda}_{-\tilde a})
+\lambda (S_a , {\tilde S}_{\tilde a})$ ($\lambda$; Grassmann odd). 
It is expressed by the mapping 
$f : \lambda \mapsto \lambda (S_a, {\tilde S}_{\tilde a})$. 
The other conditions are interpreted as kernel of the 
mapping $g$ for $(\lambda_{-a},{\tilde \lambda}_{- \tilde a})$, 
$$
g : (\lambda_{-a},{\tilde \lambda}_{- {\tilde a}}) \mapsto 
\biggl(\sum_a{\partial J^a\over \partial \phi^I}\lambda_{-a}
+\sum_{\tilde a}
{\partial \tilde{J}^{\tilde a}\over \partial \phi^I}{\tilde \lambda}_{-\tilde a},
\sum_a{\partial J^a\over \partial \xi^A}\lambda_{-a}
+\sum_{\tilde a}
{\partial \tilde{J}^{\tilde a}\over \partial \xi^A}{\tilde \lambda}_{-\tilde a}\biggr) .
$$
Then the massless fields in the left-movers are 
described by \LSconst{b,c}\ up to the gauge equivalence, 
in other words, those massless fields are represented by kernel of $g$ 
up to image of $f$.

\newsec{Conclusions}

We have studied the gauged linear sigma model on the supermanifold 
in which the fermionic chiral multiplets are introduced for the 
Grassmann odd coordinates on the target supermanifold. 

Firstly, for a simple example, we have constructed 
the $\CN=(2,2)$ supersymmetric $U(1)$ gauged linear sigma model and 
investigated its vacuum structure.
On one vacuum, the target space is reduced to the hypersurface in the 
super weighted projective space 
$\Bbb{WCP}^{m-1|n}_{(Q_1,Q_2,\cdots,Q_m|q_1,q_2,\cdots,q_n)}$.
On another vacuum, the theory becomes the bosonic 
Landau-Ginzburg orbifold, which is the same as the Landau-Ginzburg orbifold 
derived from an ordinary bosonic $\Bbb{CP}$ model. 
On the other vacuum, the fermionic Landau-Ginzburg orbifold appears 
and it is the new interesting feature on account of considering 
the supermanifold. As we have seen so far, there might be 
the transition between the bosonic and the fermionic manifolds. 

In order to study the quantum property of vacua, 
we have calculated one-loop correction to the 
expectation value of the D-field. 
By this analysis, information on 
renormalization property of the FI parameter $r$ can be evaluated. 
We obtained the condition for the one-loop divergence 
of this D-field to vanish. 
The result is consistent with the Ricci flatness condition of 
the supermanifold. If this condition is not satisfied, then  
the FI parameter $r$ runs according to the scale parameter $\mu$ and 
the theory depends on the $\mu$.

Next we have calculated the supersymmetric transformation of the Lagrangian and 
have considered two types of supersymmetric boundary conditions. 
The A-type boundary condition is realised as the brane with 
a half of the bosonic and the fermionic dimensions of the target supermanifold. 
It is the analogue of an A-brane on the Lagrangian submanifold in the 
ordinary Calabi-Yau manifold. On the other hand, 
we have shown that the B-type boundary condition 
corresponds to the subsupermanifold whose bosonic and fermionic parts 
have real even dimensions. That should be 
the analogue of a B-brane wrapped on a holomorphic cycle 
in a purely bosonic manifold.
Though Lagrangian submanifolds and holomorphic cycles in supermanifolds 
might be not understood so clearly, 
we have been able to obtain the D-branes wrapped 
on the supermanifold from the above consideration. 

In section 5, we have constructed the non-abelian gauged linear sigma model as 
an extension of the abelian case.
In order to study the structure of vacua, 
we wrote down the potential term of the scalar fields 
and analysed the D-flatness condition. 
The D-field corresponds to the moment map associated with the 
gauge symmetry group, and the structure of vacua is
determined by the vanishing locus of this field. 
For example, if we take the model with $U(m)$ flavour symmetry and 
put all the fermionic coordinates to equal zero,  
the corresponding classical vacuum becomes the well-known Grassmannian manifold.
In order to analyse the quantum property of vacua, 
we performed one-loop calculation of the D-field in terms of 
the free field approximation. 
The resulting theory generally has mass gap. 
But some condition is met, there could be scale invariant theory 
at least in this approximation.

For the purpose of construction of another supersymmetric model, 
we also have built the $(0,2)$ supersymmetric model. 
The reduction of the $(2,2)$ supersymmetric model 
has been expressed as an concrete example of this $(0,2)$ model.
We also have discussed the massless fields in this model 
and expected that they play important roles in the low energy physics.

\bigbreak\bigskip\bigskip
\centerline{{\bf Acknowledgements}}\nobreak

S.S. is grateful to M.~Fukuma, T.~Hirohashi, H.~Irie and A.~Miwa 
for useful discussions. 
The research of S.S. was supported in part by the Grant-in-Aid 
for Scientific Research (\#16740159) from the Ministry 
of Education, Culture, Sports, Science and Technology (MEXT) of Japan 
and also by the Grant-in-Aid for the 21st Century COE 
``Center for Diversity and Universality in Physics'' from MEXT.
K.S. was supported in part by the Grant-in-Aid for Scientific Research (\#14740115) from MEXT.

\appendix{A}{The boundary terms of supersymmetric transformations}

The boundary terms with the auxiliary fields 
by the supersymmetric transformation are calculated in terms of 
\STvec{}, \STchib{}, \STchiba{}, \STchif{}\ and \STchifa{}, and 
are described as follows:
$$
\delta L = -\int_{\partial\Sigma} dx^0 
(\epsilon_+ \CL^+ - \epsilon_- \CL^- 
- \Bepsilon_+ \BCL^+ + \Bepsilon_- \BCL^-), 
$$
where $\CL^+$ and $\CL^-$ are given by 
\eqn\aSTLp{\eqalign{
\CL^+ = &-\sum_I\biggl({1 \over 2\sqrt{2}}\psi^I_-(\BD_0+\BD_1)\Bphi^I
   - {1 \over 2\sqrt{2}}\Bphi^I(D_0+D_1)\psi^I_- \cr
   &- {i \over \sqrt{2}}\Bpsi^I_+ F^I
   - iQ_I\psi^I_+\sigma\Bphi^I + Q_I\Blambda_+\Bphi^I\phi^I\biggr) \cr
   &-\sum_A\biggl(-{1 \over 2\sqrt{2}}b^A_-(\BD_0+\BD_1)\Bxi^A 
   + {1 \over 2\sqrt{2}}\Bxi^A(D_0+D_1)b^A_- \cr
   &+ {i \over \sqrt{2}}\Bb^A_+\chi^A
   + iq_Ab^A_+\sigma\Bxi^A + q_A\Blambda_+\Bxi^A\xi^A\biggr) \cr
   &-{1 \over e^2}\biggl( -{i \over 2\sqrt{2}}\Blambda_-(\partial_0+\partial_1)\sigma
   +{i \over 2\sqrt{2}}\sigma(\partial_0+\partial_1)\Blambda_- 
   + {1 \over 2}(D + iv_{01})\Blambda_+ \biggr) \cr
   &-\biggl(-r + i {\theta \over 2\pi}\biggr)\Blambda_+ 
   - i\sqrt{2} \biggl(\sum_I\Bpsi_+^I{\partial \BW(\Bphi,\Bxi) \over \partial \Bphi^I}
   -\sum_A\Bb_+^A{\partial \BW(\Bphi,\Bxi) \over \partial \Bxi^A}\biggr) ,
}}
\eqn\aSTLm{\eqalign{
\CL^- = &-\sum_I\biggl(-{1 \over 2\sqrt{2}}\psi^I_+(\BD_0-\BD_1)\Bphi^I 
   + {1 \over 2 \sqrt{2}}\Bphi^I (D_0-D_1)\psi^I_+ \cr
   &- {i \over \sqrt{2}}\Bpsi^I_- F^I
   + iQ_I\psi^I_-\Bsigma\Bphi^I + Q_I\Blambda_-\Bphi^I\phi^I \biggr) \cr
   &-\sum_A\biggl({1 \over 2\sqrt{2}}b^A_+(\BD_0-\BD_1)\Bxi^A 
   - {1 \over 2\sqrt{2}}\Bxi^A(D_0-D_1)b^A_+ \cr
   &+ {i \over \sqrt{2}}\Bb^A_-\chi^A
   - iq_Ab^A_-\Bsigma\Bxi^A + q_A\Blambda_-\Bxi^A\xi^A\biggr) \cr
   &- {1 \over e^2}\biggl(-{i \over 2\sqrt{2}}\Blambda_+(\partial_0-\partial_1)\Bsigma
   +{i \over 2\sqrt{2}}\Bsigma(\partial_0-\partial_1)\Blambda_+ 
   + {1 \over 2}(D - iv_{01})\Blambda_- \biggr) \cr
   &- \biggl(-r - i {\theta \over 2\pi}\biggr)\Blambda_-
   - i\sqrt{2} \biggl(\sum_I\Bpsi_-^I{\partial \BW(\Bphi,\Bxi) \over \partial \Bphi^I}
   - \sum_A\Bb_-^A{\partial \BW(\Bphi,\Bxi) \over \partial \Bxi^A}\biggr) ,
}}
and $\BCL^+$ and $\BCL^-$ are hermitian conjugates of $\CL^+$ and $\CL^-$ 
respectively.

\listrefs
\bye